\documentclass[fleqn,usenatbib]{mnras}

\usepackage{newtxtext,newtxmath}

\usepackage[T1]{fontenc}

\DeclareRobustCommand{\VAN}[3]{#2}
\let\VANthebibliography\thebibliography
\def\thebibliography{\DeclareRobustCommand{\VAN}[3]{##3}\VANthebibliography}

\usepackage{graphicx}	
\usepackage{amsmath}	
\usepackage{booktabs}

\newcommand{\kms}{km~s$^{-1}$}

\newcommand{\teff}{$T_{\rm eff}$}
\newcommand{\msun}{M$_\odot$}

\newcommand{\msunyr}{M$_\odot$yr$^{-1}$}
\newcommand{\vsini}{$v\sin i$}

\newcommand{\vrad}{$V_{\rm r}$}
\newcommand{\rstar}{$R_\star$}

\title[Accretion and magnetism on young binaries]{Accretion and magnetism on young eccentric binaries: DQ Tau and AK Sco}

\author[K. Pouilly et al.]{
Kim Pouilly$^{1}$\thanks{New e-mail: Kim.Pouilly@unige.ch},
Axel Hahlin$^{1}$,
Oleg Kochukhov$^{1}$,
Julien Morin$^{2}$,
and Ágnes Kóspál$^{3,4,5,6}$
\\
$^{1}$Department of Physics and Astronomy, Uppsala University, Box 516, SE-75120 Uppsala, Sweden\\
$^{2}$LUPM, Université de Montpellier \& CNRS, Montpellier, Cedex 05, France\\
$^{3}$Konkoly Observatory, HUN-REN Research Centre for Astronomy and Earth Sciences, Konkoly-Thege Mikl\'os \'ut 15-17, 1121 Budapest, Hungary\\
$^{4}$Max Planck Institute for Astronomy, K\"{o}nigstuhl 17, 69117 Heidelberg, Germany\\
$^{5}$ELTE E\"{o}tv\"{o}s Lor\'{a}nd University, Institute of Physics, P\'{a}zm\'{a}ny P\'{e}ter s\'{e}t\'{a}ny 1/A, 1117 Budapest, Hungary\\
$^{6}$CSFK, MTA Centre of Excellence, Konkoly-Thege Mikl\'os \'ut 15-17, 1121 Budapest, Hungary
}

\date{Accepted 2024 February 01. Received 2024 February 01; in original form 2023 August 21}

\pubyear{2024}

\begin{document}
\label{firstpage}
\pagerange{\pageref{firstpage}--\pageref{lastpage}}
\maketitle

\begin{abstract}
The accretion and ejection of mass in pre-main sequence (PMS) stars are key processes in stellar evolution as they shape the stellar angular momentum transport necessary for the stars’ stability.
Magnetospheric accretion onto classical T Tauri stars and low-mass PMS stars has been widely studied in the single-star case.
This process can not be directly transferred to PMS binary systems, as tidal and gravitation effects, and/or accretion from a circumbinary disc (with variable separation of the components in the case of eccentric orbits) are in place.
This work examines the accretion process of two PMS eccentric binaries, DQ Tau and AK Sco, using high-resolution spectropolarimetric time series.
We investigate how magnetospheric accretion can be applied to these systems by studying the accretion-related emission lines and the magnetic field of each system.
We discover that both systems are showing signs of magnetospheric accretion, despite their slightly different configurations, and the weak magnetic field of AK Sco.
Furthermore, the magnetic topology of DQ Tau A shows a change relative to the previous orbital cycle studied: previously dominated by the poloidal component, it is now dominated by the toroidal component.
We also report an increase of the component’s accretion and the absence of an accretion burst at the apastron, suggesting that the component’s magnetic variation might be the cause of the inter-cycle variations of the system’s accretion.
We conclude on the presence of magnetospheric accretion for both systems, together with gravitational effects, especially for AK Sco, composed of more massive components.
\end{abstract}




\section{Introduction}

Understanding the accretion of pre-main sequence (PMS) stars is a major aim of stellar and planetary formation and evolution studies, because, together with the ejections processes, it governs the transport of angular momentum which ensures the star's stability, and shapes the disc, birthplace of exoplanets.
The accretion process of low-mass PMS objects known as classical T Tauri stars (cTTSs), where accreting material from their accretion disk is driven by their magnetic field, has been widely studied on single stars \citep[e.g.,][for recent studies]{Alencar18, Donati19, Bouvier20a, Bouvier20b, Pouilly20, Pouilly21}.
The magnetic field lines connected to the disc exert a magnetic pressure pulling the material out off the disc plane and forcing an accretion through accretion funnel flows.
The different features of this magnetospheric accretion process, such as the accretion funnel flow itself, the accretion shock it produces at the stellar surface, the magnetic field-disc connection, or directly the stellar magnetic field, can be studied using spectroscopic and spectropolarimetric time series, mapping the stellar surface and tracing the accretion signatures over several stellar rotation cycles.

However, this scheme cannot be directly applied to close PMS binaries, because of the presence of a companion implying tidal and gravitational interactions, and/or of the accretion from a circumbinary disc instead of a circumstellar disc. 
Only a few of these systems have been studied using this methodology \citep[V4046 Sgr, V1878 Ori, and DQ Tau in][respectively]{Donati11, Lavail20, Pouilly23}, and no clear common accretion scheme (if there is one) has been established. 
This work aims to study and compare the accretion of two equal-mass PMS eccentric binary systems: DQ Tau and AK Sco.
These two systems are particularly suitable for such a study because of their orbital period (15.8 and 13.6 d for DQ Tau and AK Sco, respectively), allowing them to be fully covered by spectropolarimetric time series, and thus obtain observation at each phase of the orbital cycle.
Furthermore, both systems are spectroscopic binaries, composed of equal-mass accreting PMS star components, orbiting on an eccentric orbit, and surrounded by a circumbinary disc, which allows us to directly compare the two accretion processes.

DQ Tau (RA 04$^{\rm h}$ 46$^{\rm min}$ 53$^{\rm s}$.058, Dec +17$^{\circ}$ 00' 00''.14), consist of two $\sim$0.6 \msun\ (M0-type) cTTSs orbiting with a e$\sim$0.6 eccentricity.
The separation at the periastron of the system, 12.5 \rstar, results in an interaction between the two magnetospheres \citep{Salter10, Getman11}, which affects the components' accretion.
\cite{Tofflemire17} revealed, from multi-band photometry, an enhancement by an order of magnitude of the mass accretion at periastron, called the "pulsed accretion" phenomenon \citep{Artymowicz96, Mathieu97}.
This phenomenon was also studied from the K2 light curve by \cite{Kospal18}, revealing strong bursting events at each periastron passage, but sometimes at the apastron as well.
Each bursting event is separated by a very stable sinusoidal modulation, typical of a spotted surface, yielding a precise derivation of the stars' rotation period (P = 3.017 d).
In addition, the authors determined an inner radius of the circumbinary disc of 0.13 AU, but later \cite{Muzerolle19} drew a more complex picture with an orbital phase-dependent location and geometry of the emitting material, with an inner disc radius closer to 0.28 AU.
More recently, two studies focused on this system's accretion.
\cite{Fiorellino22}, from X-SHOOTER spectra at 8 different epochs, detected the pulsed accretion as well and showed that the two components are accreting material, with the main accretor changing between the two components.
Finally, \cite{Pouilly23} (hereafter Paper I) performed the analysis of a high-resolution spectropolarimetric time series covering a complete orbital cycle with a $\sim$1-day sampling. 
This study confirmed the previous works' conclusion and revealed accretion signatures compatible with the single-like magnetospheric accretion process obtained, as well as an apsidal motion, a precession of the orbit, never observed in such systems before, and hypothesised to originate from the interactions at the periastron and/or at the apastron.
Paper I also provided the first-ever magnetic field study of this system, deriving similar small-scale field strength for both components, but significantly different large-scale magnetic topology. 
Both topologies are compatible with the magnetospheric accretion scheme, mainly poloidal and dominated by the dipole component, but the mean global magnetic field strength of the primary is much weaker (160 G) than the secondary (570G).

AK Sco (RA 16$^{\rm h}$ 54$^{\rm min}$ 44$^{\rm s}$.849, Dec -36$^{\circ}$ 53' 18''.57), is PMS spectroscopic binary on an eccentric orbit with a close separation at periastron as well (11 \rstar), composed of two 1.3\msun\ F5-type young stellar objects, and a circumstellar disc with a larger inner radius of 0.4 AU \citep{Alencar03}.
The Table~\ref{tab:parameters} provides a summary of relevant parameters for the two systems to allow the reader a direct comparison.
The accretion of this system was studied in detail by \cite{Alencar03} using high-resolution spectroscopy obtained between 1998 and 2000.
The authors showed accretion-ejection processes governed by the orbital motion, and even if the two components are accreting material, they conclude from a fit of the spectral energy distribution that a substantial gap is present in the circumbinary disc, filled with gas that is powering the accretion without any accretion enhancement at periastron as expected by the models and seen on DQ Tau \citep[see numerical simulations by][]{Gunther02}.
They also revealed the presence of small-scale structures inside the binary orbit, producing a phase-dependent obscuration of the components.
The way AK Sco's components are pulling material from such a distant disc resides in its binarity and was revealed by \cite{GomezDeCastro13}.
Indeed, the variable gravitational potential efficiently drags the material from the disc's inner border at the apastron, producing spiral waves within the inner disc and forming ring-like structures around each component.
At periastron, these structures get in contact, losing angular moment and producing accretion bursts.
Later, \cite{GomezDeCastro20} noticed a cycle-to-cycle variation of the accretion, as well as the expected enhancement at the periastron, using a Hubble monitoring during three consecutive periastron passages. 
The role of the magnetic field in this scenario is still unclear, and so the link with the single-like magnetospheric accretion, as well as a common scheme of eccentric binaries accretion (if there is one) remains unknown.
The magnetism of AK Sco has been studied by \cite{Jarvinen18}, from six HARPSpol observations.
These authors detected a $\sim -$80 G longitudinal magnetic field on the secondary, which is weak but does not exclude the possibility of a magnetically-driven accretion.

\begin{table}
    \centering
        \caption{Parameters of DQ Tau and AK Sco systems from the literature}
    \begin{tabular}{l c c}
        \hline
         & DQ Tau & AK Sco \\ 
        \hline
        Components SpT & M0 & F5  \\
        Components mass (\msun) & 0.6 & 1.3 \\
        $P_{\rm orb}$ (d) & 15.8 & 13.6  \\
        Eccentricity & 0.6 & 0.5 \\
        Periastron separation (AU) & 12.5 & 11  \\
        Inner disc radius (AU) & 0.13-0.28 & 0.4  \\    
    \hline
    \end{tabular}
    \label{tab:parameters}
\end{table}

In this work, we analysed the two systems to compare their behaviour regarding their accretion process. 
The DQ Tau data set consists of 11 new high-resolution spectropolarimetric spectra obtained in late 2022, and we re-analysed the AK Sco data set from 2016-2017 studied by \citep{Jarvinen18}, in addition to eight new HARPSpol spectra obtained in 2019 and 2022.
This paper is organised as follows. A description of the observations is provided in Sect.~\ref{sec:obs}, the results as presented in Sect.~\ref{sec:results}, beginning with DQ Tau and followed by AK Sco. 
We discuss the results on each system separately and compare them in Sect.~\ref{sec:discuss}, before concluding this paper in Sect.~\ref{sec:ccl}.

\section{Observations}
\label{sec:obs}

The spectropolarimetric observations were acquired using two different spectropolarimeters for the two objects.
We describe both data sets here.

DQ Tau was observed using the Echelle SpectroPolarimetric Device for the Observation of Stars \citep[ESPaDOnS,][]{Donati03} at the Canada-France-Hawaii telescope (CFHT), which covers the 370 to 1050 nm wavelength range and reaches a resolving power of 68~000. 
Between 2022 September 11 and 2022 October 20, we obtained 11 observations, themselves composed of 4 sub-exposures in different polarimeter configurations, allowing us to derive the intensity (Stokes I), the circularly polarised (Stokes V) and the Null spectra.
These observations were reduced using the \texttt{Libre-ESpRIT} package \citep{Donati97}, and reach the signal-to-noise ratio (S/N) between 107 and 139 for the Stokes I, and between 98 and 128 for the Stokes V.
The log of these observations is summarised in Table~\ref{tab:logObs}.

\begin{table}
    \centering
        \caption{Log of ESPaDOnS observations of DQ Tau. The columns are listing the calendar and heliocentric Julian dates of observations, the S/N for the spectral pixel at the order 31 (731 nm) for the Stokes I and V spectra, the effective S/N of the LSD Stokes V profiles (see Sect.~\ref{subsubsec:LSD}), and the orbital phases computed from the orbital elements derived in this work (Sect.~\ref{subsubsec:LSD}).}
    \begin{tabular}{l l l l l l  }
        \hline
        \hline
        Date & HJD & S/N$_{\rm I}$ & S/N$_{\rm V}$ & S/N$_{\rm LSD}$ & $\phi_{\rm orb}$ \\ 
        (2022) & ($-$2~450~000 d) & & & &  \\
        \hline
        11 Sep & 9834.04246 & 120 & 108 & 5159 & 0.77 \\
        12 Sep & 9835.00880 & 108 & 98 & 4521 & 0.83 \\
        14 Sep & 9837.04838 & 125 & 113 & 5400 & 0.96 \\
        15 Sep & 9838.02501 & 107 & 99 & 4303 & 0.02 \\
        07 Oct & 9860.10318 & 132 & 117 & 5959 & 0.42 \\
        15 Oct & 9868.10419 & 133 & 120 & 5929 & 0.92 \\
        16 Oct & 9868.98051 & 139 & 128 & 6222 & 0.98 \\
        17 Oct & 9869.95300 & 130 & 121 & 5648 & 0.04 \\
        18 Oct & 9871.02597 & 125 & 114 & 5620 & 0.11 \\
        19 Oct & 9872.02990 & 126 & 116 & 5774 & 0.17 \\
        20 Oct & 9872.94421 & 123 & 114 & 5425 & 0.23  \\

    \hline
    \end{tabular}
    \label{tab:logObs}
\end{table}

The observations of AK Sco were performed using the High Accuracy Radial velocity Planet Searcher, in polarimetric mode \citep[HARPSpol,][]{Snik08} mounted at the 3.6 m telescope of the European Southern Observatory (ESO), and obtained from the ESO HARPS archive website.
The resolving power of this instrument is about 110~000 covering the 378$-$691 nm spectral range.
This data set consists of 19 observations split into four sub-sets over four years (2016, 2017, 2019, and 2022) and was reduced using the \texttt{REDUCE} pipeline \citep{Piskunov02}.
However, only 8 of these observations are composed of 4 sub-exposures\footnote{2016 June 15, 2017 June 04, 2017 June 04, 2022 April 27, 2022 April 29, 2022 April 30, and 2022 May 01}.
The 2-sub-exposures configuration of the other observations allows the derivation of the Stokes I and V spectra, but no Null can be computed.
The log of these observations is presented in Table~\ref{tab:logObsAK}.

\begin{table}
    \centering
        \caption{Journal of the HARPSpol observations of AK Sco. The columns consist of the calendar and heliocentric Julian dates of observations, the S/N for the spectral pixel at $\lambda \approx$ 520 nm for I and V spectra, the effective S/N of the LSD Stokes V profiles (see Sect.~\ref{subsubsec:LSDAK}), and the orbital phases computed from the orbital elements derived in this work (Sect.~\ref{subsubsec:LSDAK}).}
    \begin{tabular}{l l l l l l l}
        \hline
        \hline
        Date & HJD & S/N$_{\rm I}$ & S/N$_{\rm V}$ & S/N$_{\rm LSD}$ & $\phi_{\rm orb}$\\ 
         & ($-$2~450~000 d) & & & & \\
        \hline
        15 Jun 2016 & 7554.79823 & 144 & 84 & 10186 & 0.94 \\
        16 Jun 2016 & 7555.78259 & 152 & 89 & 10957 & 0.01 \\
        04 Jun 2017 & 7908.70690 & 202 & 118 & 14475 & 0.95 \\
        05 Jun 2017 & 7909.71661 & 263 & 152 & 18832 & 0.02 \\
        06 Jun 2017 & 7910.60562 & 181 & 107 & 13169 & 0.09 \\
        07 Jun 2017 & 7911.68494 & 175 & 101 & 12784 & 0.17 \\
        14 Jun 2019 & 8648.75645 & 133 & 78 & 9197 & 0.33 \\
        15 Jun 2019 & 8649.63905 & 109 & 63 & 7668 & 0.39 \\
        16 Jun 2019 & 8650.63415 & 112 & 63 & 7910 & 0.46 \\
        17 Jun 2019 & 8651.67538 & 101 & 59 & 7089 & 0.54 \\
        18 Jun 2019 & 8652.71917 & 63 & 36 & 4324 & 0.62 \\
        24 Apr 2022 & 9693.83067 & 151 & 86 & 13518 & 0.12 \\
        25 Apr 2022 & 9694.76515 & 108 & 61 & 9744 & 0.19 \\
        26 Apr 2022 & 9695.81443 & 92 & 53 & 8125 & 0.26 \\
        27 Apr 2022 & 9696.77257 & 95 & 55 & 8523 & 0.33 \\
        28 Apr 2022 & 9697.73313 & 79 & 46 & 7016 & 0.40 \\
        29 Apr 2022 & 9698.81650 & 153 & 88 & 13686 & 0.48 \\
        30 Apr 2022 & 9699.78042 & 214 & 124 & 19241 & 0.55 \\
        01 May 2022 & 9700.76676 & 35 & 19 & 2469 & 0.63 \\

    \hline
    \end{tabular}
    \label{tab:logObsAK}
\end{table}

\section{Results}
\label{sec:results}

In this section we present the results obtained from the datasets described is Section \ref{sec:obs}. 
The results will be split between the two objects, DQ Tau and AK Sco, and include the spectroscopic and polarimetric analysis.

\subsection{DQ Tau}
Here we present results concerning the ESPaDOnS observations of DQ Tau. 
We start with the Least Squared Deconvolution \citep[LSD - ][]{Donati97} profiles and radial velocity derivation, before the emission lines analysis, and end with the magnetic study.

\subsubsection{LSD profiles and radial velocity}
\label{subsubsec:LSD}

\begin{figure*}
    \centering
    \includegraphics[width=.49\textwidth]{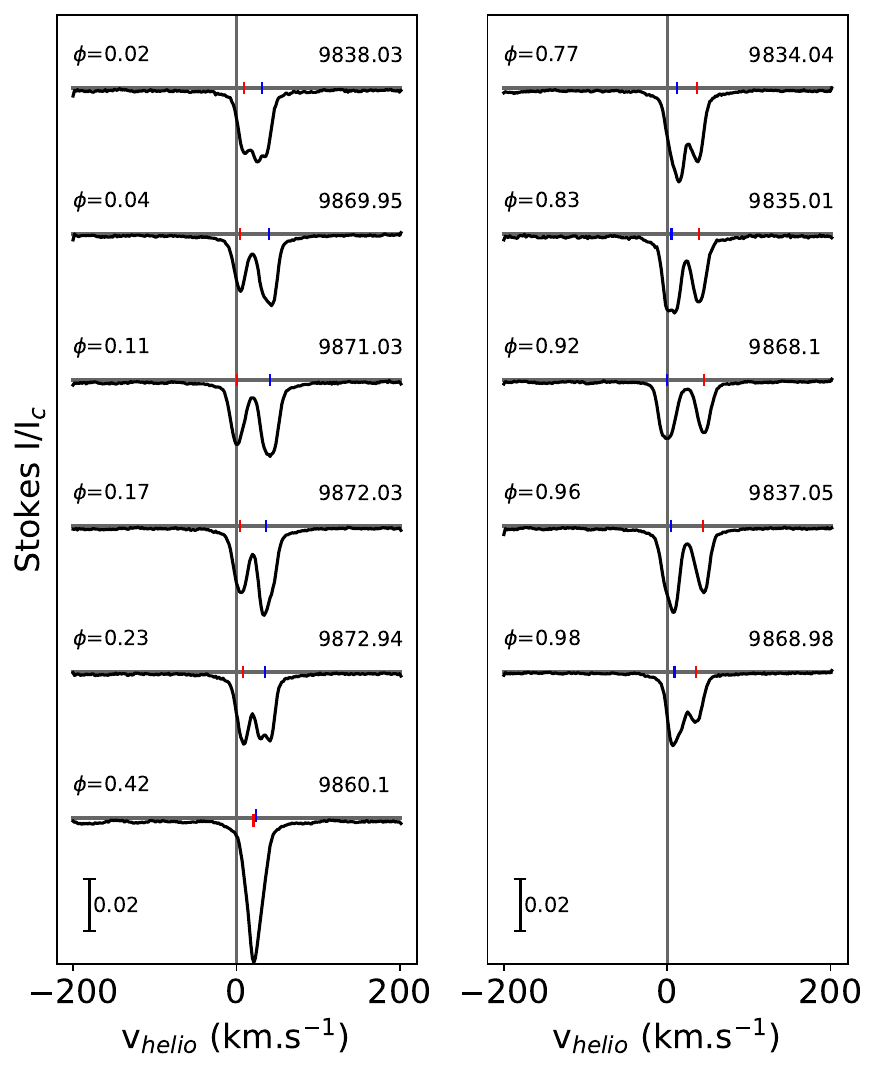}
    \includegraphics[width=.49\textwidth]{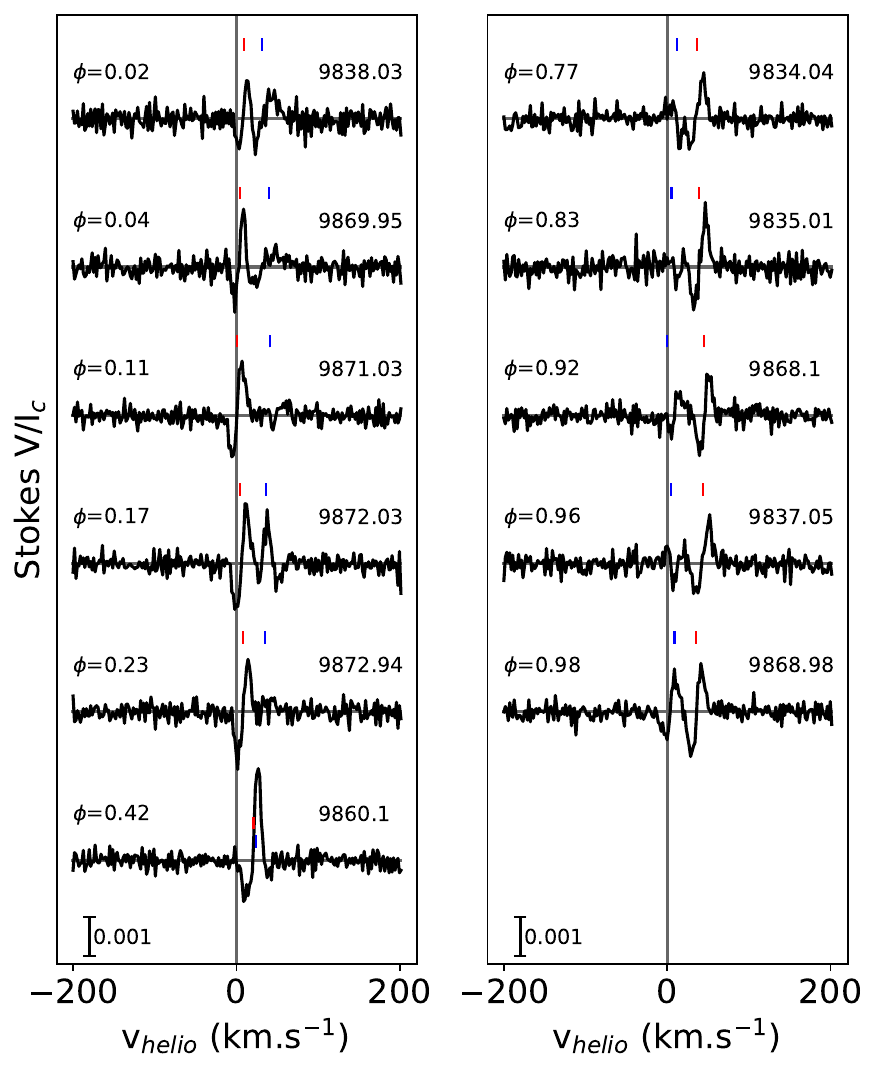}
    \caption{LSD Stokes I \textit{(left)} and V \textit{(right)} profiles of the 2022 DQ Tau ESPaDOnS dataset. The blue (red) ticks are illustrating the velocity of the A (B) component. The orbital phase and HJD ($-$2~450~000 days) are indicated on the left and right of each profile, respectively. The vertical scale is indicated at the bottom left of each plot.}
    \label{fig:lsdprof}
\end{figure*}

To derive the LSD Stokes I (unpolarised) and V (circularly polarised) profile of DQ Tau's dataset, we used the same procedure and parameters as in Paper I. This means a mean wavelength of 520 nm, a Landé factor of 1.2, an intrinsic line depth of 0.2, and the usage of a line list provided by the VALD database \citep{Ryabchikova15} for the ESPaDOnS wavelength range, from which we removed the regions containing emission lines or telluric absorption.
About 12 000 lines were taken into account, and the resulting Stokes I (V) profiles show signal-to-noise ratio (S/N) between 919 and 1062 (4303 and 6222). 
The profiles are shown in Fig.~\ref{fig:lsdprof}.
The Null profiles, that cancel the polarisation signal, are featureless and their dispersion is consistent with the uncertainties, they are thus not shown in this work.

\begin{figure}
    \centering
    \includegraphics[width=.45\textwidth]{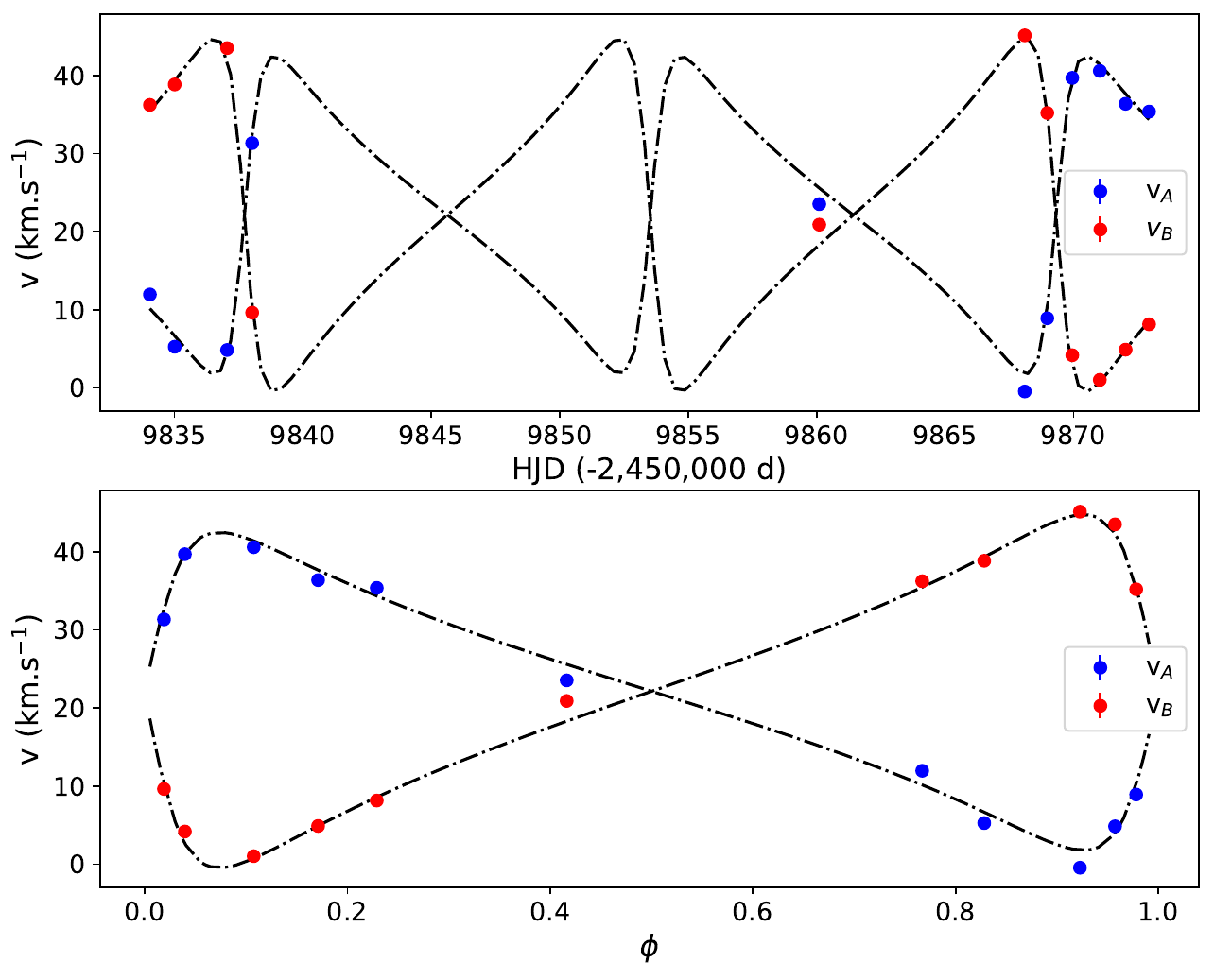}
    \caption{Radial velocities of the primary \textit{(blue)} and secondary \textit{(red)} components of DQ Tau. The dash-dotted lines is the corresponding fit of the orbital solution.}
    \label{fig:vrad}
\end{figure}

Then we derived the radial velocity of each component for each observation using the iterative disentangling procedure described in Paper I, with the orbital solution of that paper as a first guess.
The results are summarized in Table~\ref{tab:vrad} and shown in Fig.~\ref{fig:vrad}.

\begin{table}
    \centering
        \caption{Radial velocities of DQ Tau's A-B pair and their uncertainties computed from the LSD Stokes I disentangling procedure.}
    \begin{tabular}{l | l l | l l}
        \hline
        HJD & \vrad(A) & $\delta$\vrad(A) & \vrad(B) & $\delta$\vrad(B) \\
        ($-$2~450~000 d) & \kms & \kms & \kms & \kms \\
        \hline
        9834.04246 & 11.96 & 0.64 & 36.26 & 0.49 \\
        9835.00880 & 5.26 & 0.54 & 38.87 & 0.38 \\
        9837.04838 & 4.84 & 0.54 & 43.54 & 0.37 \\
        9838.02501 & 31.36 & 0.29 & 9.63 & 0.19 \\
        9860.10318 & 23.55 & 0.60 & 20.91 & 0.46 \\
        9868.10419 & -0.47 & 0.20 & 45.17 & 0.40 \\
        9868.98051 & 8.91 & 0.29 & 35.22 & 0.30 \\
        9869.95300 & 39.72 & 0.32 & 4.18 & 0.16 \\
        9871.02597 & 40.61 & 0.44 & 1.01 & 0.28 \\
        9872.02990 & 36.40 & 0.58 & 4.88 & 0.41 \\
        9872.94421 & 35.40 & 0.36 & 8.14 & 0.32 \\

    \hline
    \end{tabular}
    \label{tab:vrad}
\end{table}

Finally, we fitted an orbital solution using a Levenberg-Marquart algorithm (LMA) on these radial velocity values to extract the orbital elements and compare them with the results of Paper I.
We present the results in Table~\ref{tab:orbElements}, and the corresponding curves are shown in Fig.~\ref{fig:vrad}.

\begin{table}
    \centering
    \caption{Orbital elements of the DQ Tau system. $P_{\rm orb}$ is the orbital period, $\gamma$ is the systemic velocity, $K_1$ (and $K_2$) is the semi-amplitude of the A (and B) component, $e$ is the eccentricity, $\omega$ is the argument at periastron, and $T_{\rm peri}$ is the time at periastron.}
    \begin{tabular}{l c c}
        \hline
        Parameter & Paper I & This work \\
        \hline
        $P_{\rm orb}$ (days)& 15.80 $\pm$ 0.01 & 15.802 $\pm$ 0.002 \\
        $\gamma$ (\kms) & 21.9 $\pm$ 0.4 & 22.1 $\pm$ 0.3 \\
        $K_1$ (\kms) & 20.5 $\pm$ 0.6 & 20.3 $\pm$ 1.1 \\
        $K_2$ (\kms) & 22.2 $\pm$ 0.4 & 22.6 $\pm$ 1.2 \\
        $e$ & 0.58 $\pm$ 0.10 & 0.59 $\pm$ 0.04 \\
        $\omega$ ($^{\circ}$) & 266.0 $\pm$ 3.6 & 267.6 $\pm$ 2.8 \\
        $T_{\rm peri}$ (HJD - 2~400~000) & 59173.73 $\pm$ 0.10 & 59821.89 $\pm$ 2.82 \\
    \hline
    \end{tabular}
    \label{tab:orbElements}
\end{table}

All the orbital parameters are perfectly consistent within the uncertainties with those from Paper I.
The apsidal motion derived in Paper I means that we should have detected a shift of 2$^\circ$ in the argument at periastron on this dataset.
The measured difference is 1.6$^\circ$, but this may be coincidental considering the $\sim$3$^\circ$ uncertainty on this parameter.

\subsubsection{Emission lines}
\label{subsubsec:EmLineDQ}

In this section we present our analysis of the main emission lines of DQ Tau's spectrum, H$\alpha$, H$\beta$, and H$\gamma$, and of the accretion tracer He~\textsc{i} line at 587.6 nm.  
As the Balmer lines are partially formed in the accretion funnel flow \citep{Muzerolle01} and the narrow component (NC) of the He~\textsc{i} line is formed in the post-shock region of the accretion shock \citep{Beristain01}, we aimed to use these lines to trace the accretion of both components of the system.

\begin{figure}
    \centering
    \includegraphics[width=.45\textwidth]{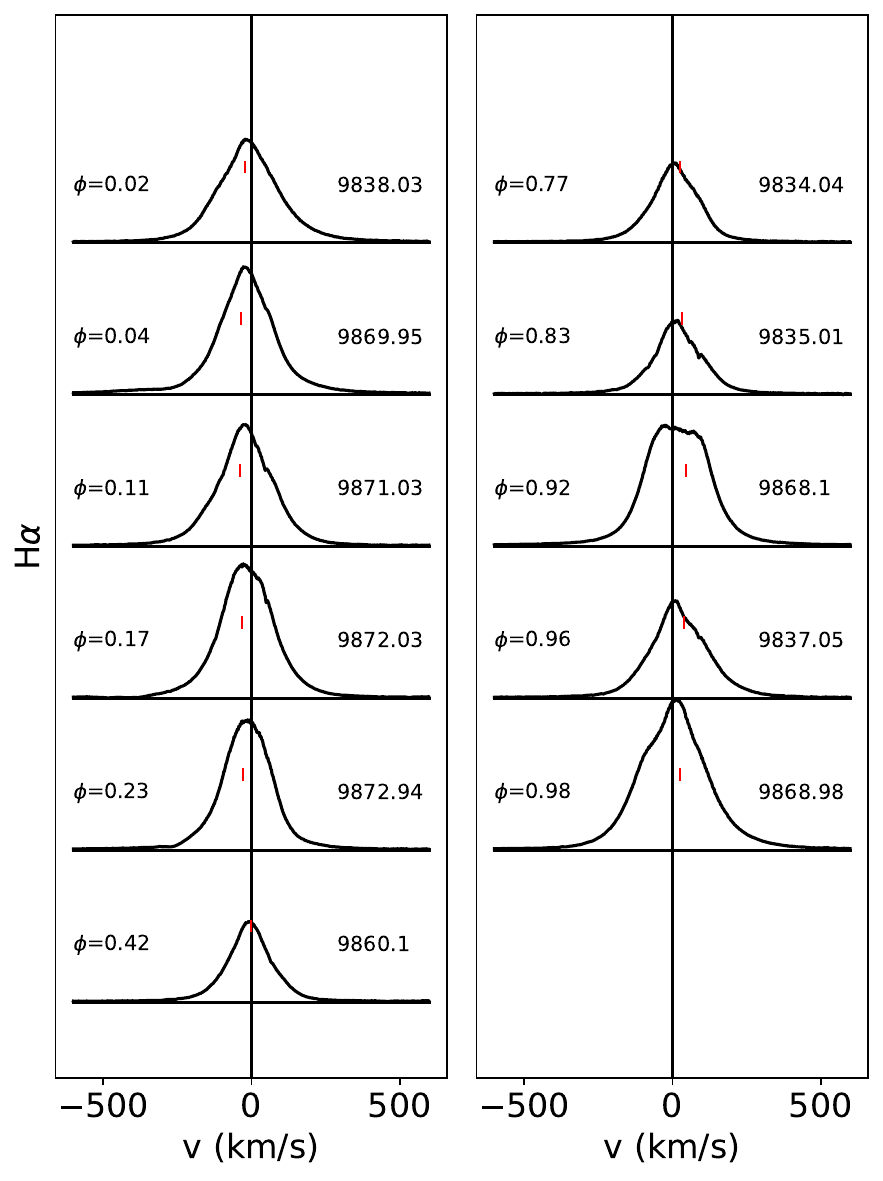}
    \caption{Residual H$\alpha$ lines of DQ Tau. These were corrected for the primary's radial velocity, the vertical line at 0 \kms\ thus corresponds to its velocity. The red ticks are indicating the velocity of the secondary in this reference frame. The orbital phase and the HJD are indicated at the left and right of each profile, respectively.}
    \label{fig:ha}
\end{figure}

The three Balmer residual lines studied, corrected from the photospheric and chromospheric components the same way as in Paper I, exhibit the same behaviour, we are thus showing only the H$\alpha$ line in Fig.~\ref{fig:ha}, the H$\beta$ and H$\gamma$ lines are available in Fig.~\ref{fig:hbhg}.
The variability of the line is modest compared to Paper I, and shows a maximum close to the periastron passage ($\phi$ = 0.98), without any maximum around the apastron.
We noticed the flat shape of the line at $\phi$ = 0.92, reminiscent of a strong emission by the two components simultaneously, as suggested by the double-peaked shape exhibited by the corresponding H$\beta$ profile.

The periodicity analysis of these lines did not reveal any periodic modulation, probably due to the sparse time sampling of this dataset. 
As information, we performed a 2D periodogram on the last six observations which have a 1-day sampling and retrieved the periodicity around 4.5 to 5 days revealed in Paper I and ascribed to the secondary's rotation period.
However, the false alarm probability (FAP) of 0.2 for this signal does not allow one to claim a robust detection here.

Finally, we computed the mass accretion rate from the H$\alpha$, H$\beta$, and H$\gamma$ lines using the following relations described in \cite{Alcala17}:

\begin{align}
    \log \left(L_{\rm acc}\right) &= a \log \left(L_{\rm line}\right) + b,
    \\
    L_{\rm line} &= 4 \pi d^{2} F_{\rm line},
    \\
    F_{\rm line} &= F_{\rm 0} \cdot EW \cdot 10^{-0.4 m_{\lambda}},
\end{align}
and,
\begin{equation}
    L_{\rm acc} = \frac{G M_{\star} \dot{M}_{\rm acc}}{R_{\star}} \left[ 1 - \frac{R_{\star}}{R_{\rm t}}\right].
\end{equation}

\begin{figure}
    \centering
    \includegraphics[width=.48\textwidth]{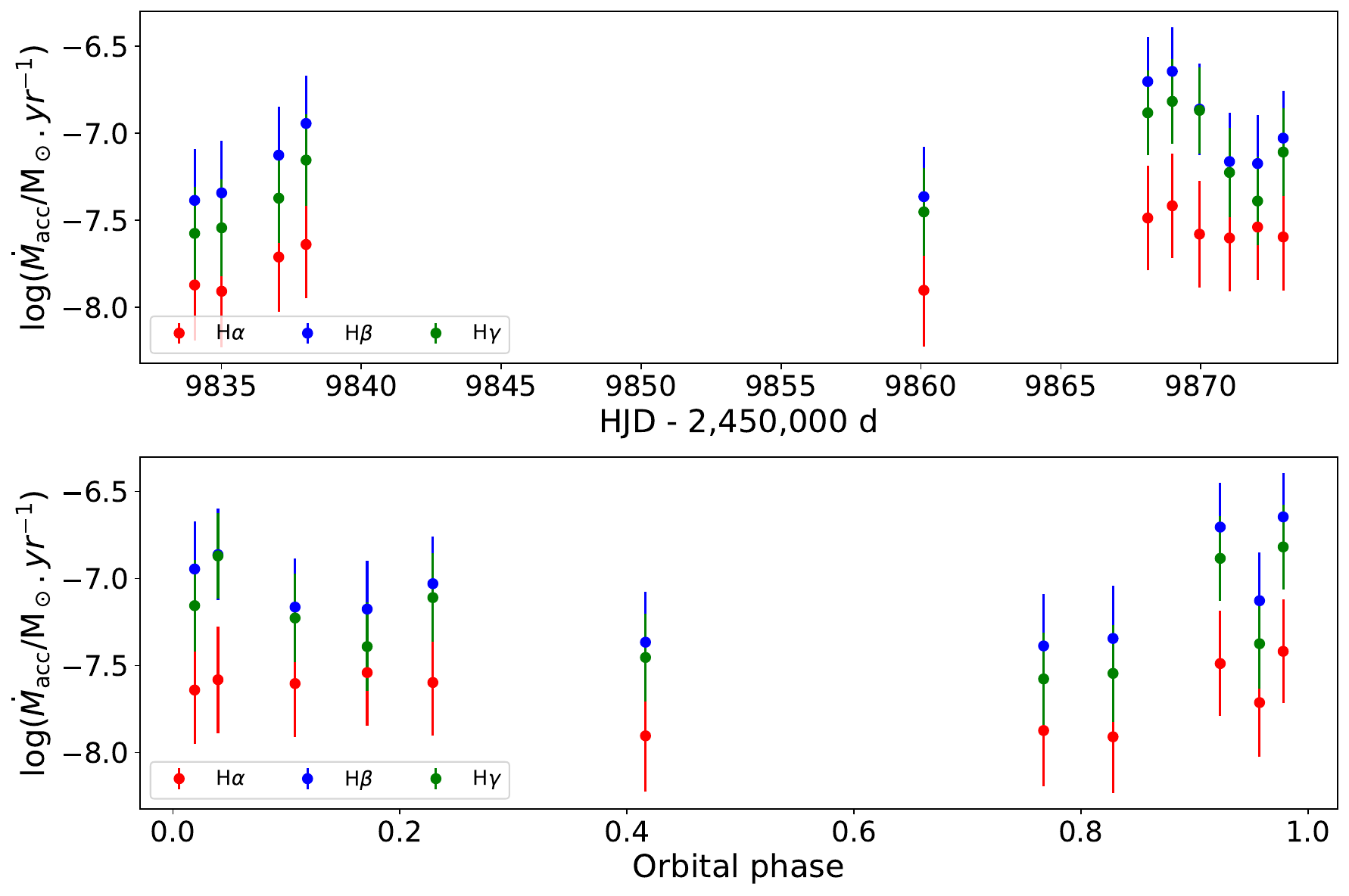}
    \caption{Mass accretion rate of DQ Tau computed from the three Balmer lines H$\alpha$ \textit{(red)}, H$\beta$ \textit{(blue)}, and H$\gamma$ \textit{(green)}.}
    \label{fig:macc}
\end{figure}

\noindent $L_{\rm acc}$ and $L_{\rm line}$ are the accretion and the line luminosity, respectively, $a$ and $b$ are the linear coefficients given by \cite{Alcala17}.
The distance $d$ is given by the Gaia EDR3 \citep{Gaia21} and $F_{\rm line}$ is the line flux, computed from $F_{\rm 0}$, the reference flux at the wavelength corresponding to the line, $EW$ is the line equivalent width, and $m_{\lambda}$ is the star's magnitude in the selected wavelength and corrected for the extinction \citep[A$_{\rm V}$=1.7 mag, ][]{Fiorellino22}.
$R_{\rm t}$ is the truncation radius, typically 5 $R_\star$ for the cTTSs \citep{Bouvier07a}. For the same reason as Paper I and \cite{Tofflemire17}, we have neglected the phase dependency of this parameter induced by the orbital motion.
The results are presented in Fig.~\ref{fig:macc}.

The values computed for each line are consistent within 2$\sigma$ between them, and the mean mass accretion rate (about 7$\times$10$^{-8}$ M$_\odot yr^{-1}$) is higher than the epochs studied in Paper I, \cite{Fiorellino22}, and \cite{Muzerolle19}.
Furthermore, as suggested by the line's variability, an enhancement is observed only at periastron (between phase 0.9$-$1.05), and the minimal value is reached near the apastron.

\begin{figure}
    \centering
    \includegraphics[width=.45\textwidth]{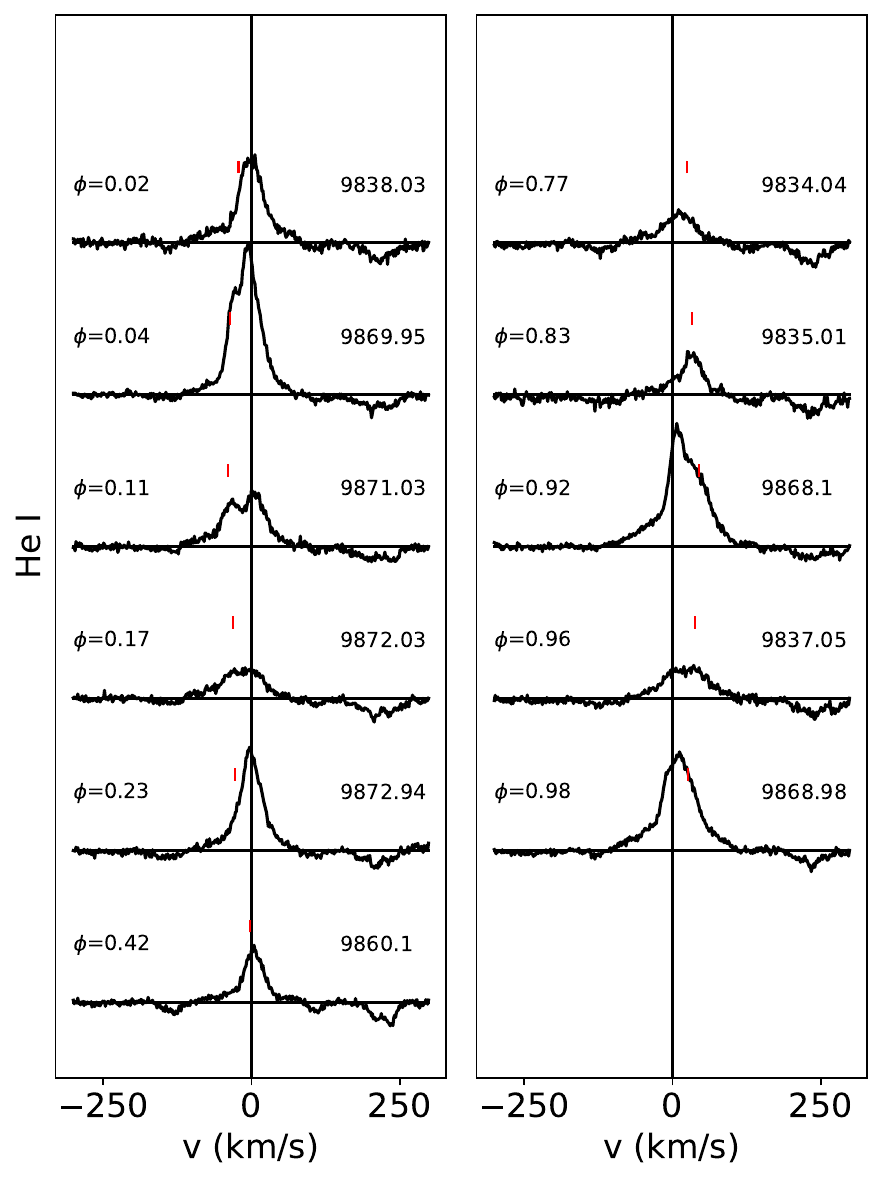}
    \caption{Same as Fig.~\ref{fig:ha} for DQ Tau's He~\textsc{I} D3 lines.}
    \label{fig:he}
\end{figure}

The He~\textsc{i} at 587.6 nm (He~\textsc{i} D3) presented in Fig.~\ref{fig:he} displays a large variability, and often an asymmetric shape of the NC.
This suggests that the accretion shocks of the two components are visible at these phases.
We expect the NC to be modulated on the rotation period of the emitting component, however, given the time sampling of our dataset, a periodogram analysis of the line did not yield a definite detection. 
As for the Balmer lines analysis, a period around 4.5 d is found using the last six observations, but the FAP (about 0.3) is too high to be considered as a reliable period.
This signal vanishes when using the whole dataset.

\begin{figure}
    \centering
    \includegraphics[width=.35\textwidth]{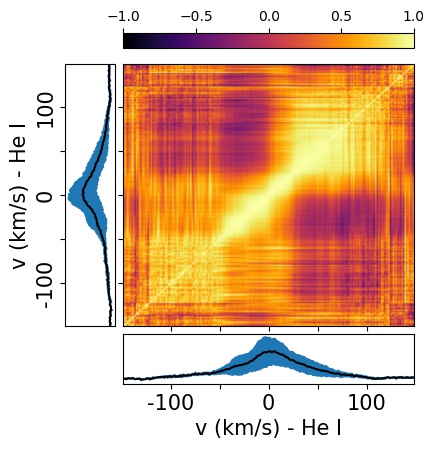}
    \caption{Autocorrelation matrix of the He~\textsc{i} D3 line for DQ Tau. The colors indicate the value of the correlation coefficient. The two panels along the x- and y-axis are showing the mean He~\textsc{i} D3 profile \textit{(black)} with its variance \textit{(blue)}.}
    \label{fig:CMhe}
\end{figure}

We derived the autocorrelation matrix of the He~\textsc{i} line, which is the derivation of a Pearson correlation coefficient between the different velocity channels of the line.
The result is shown in Fig.~\ref{fig:CMhe} and exhibits four different regions highly correlated (correlation coefficient above 0.99), in agreement with the signature of both components' accretion shock. One feature from -20 to 5 \kms, which corresponds to the NC of the primary, and three others from -40 to -20, 5 to 25, and 30 to 60 \kms, which are the locations of the second NC ascribed to the secondary's emission.

\subsubsection{Zeeman-Doppler Imaging}
\label{subsec:ZDI}

We performed a Zeeman-Doppler Imaging (ZDI) analysis to derive the large-scale magnetic field topology of DQ Tau and compare it to the previous maps of Paper I. 
The \texttt{inversLSDB} \citep{Rosen18} implementation of the ZDI code by \cite{Kochukhov14} was used on the LSD profiles shown in Fig.~\ref{fig:lsdprof}.
We used the same local profiles and stellar parameters as in Paper I and the orbital solution from Table~\ref{tab:orbElements}.

\begin{figure*}
    \centering
    \includegraphics[width=.45\textwidth]{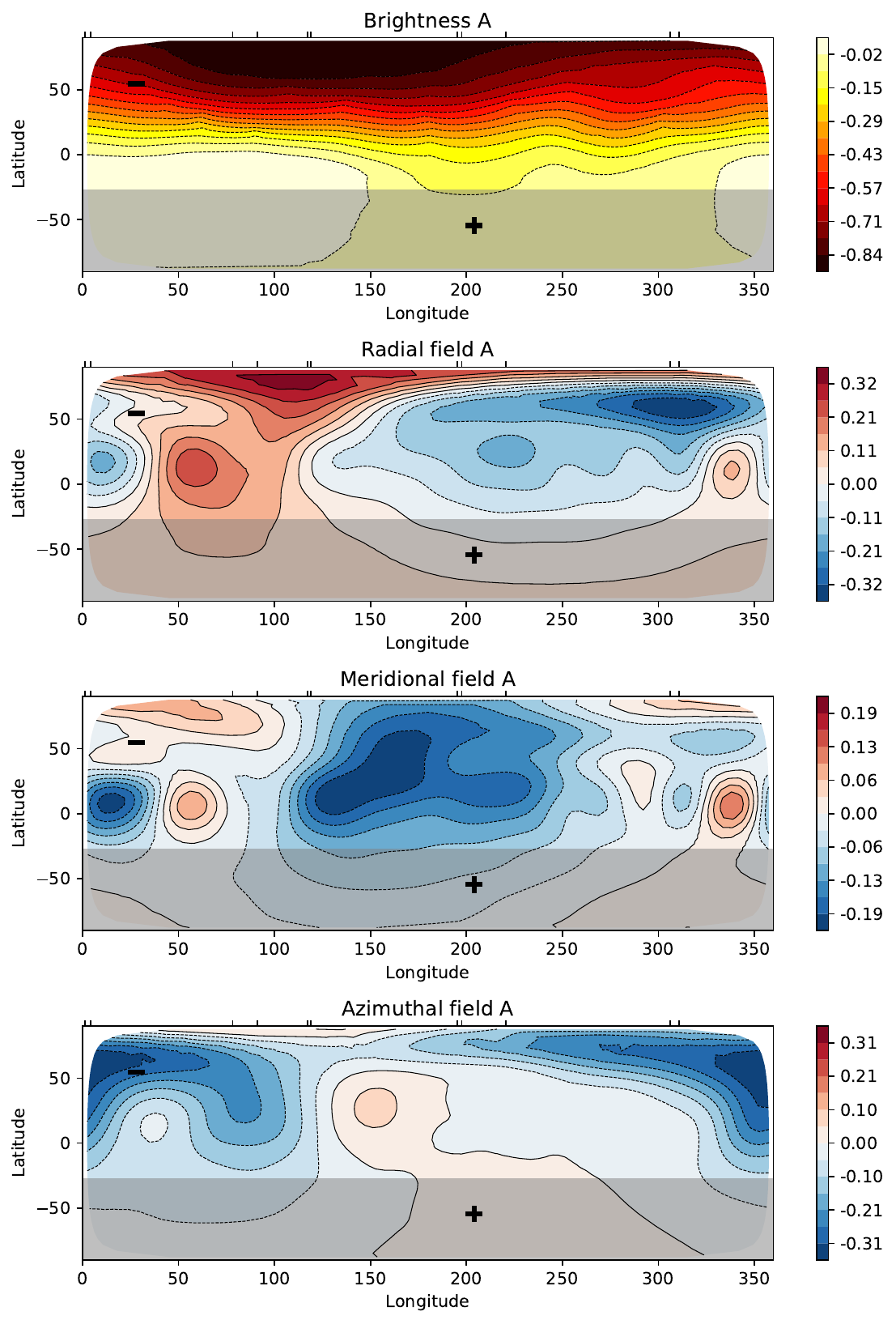}
    \includegraphics[width=.45\textwidth]{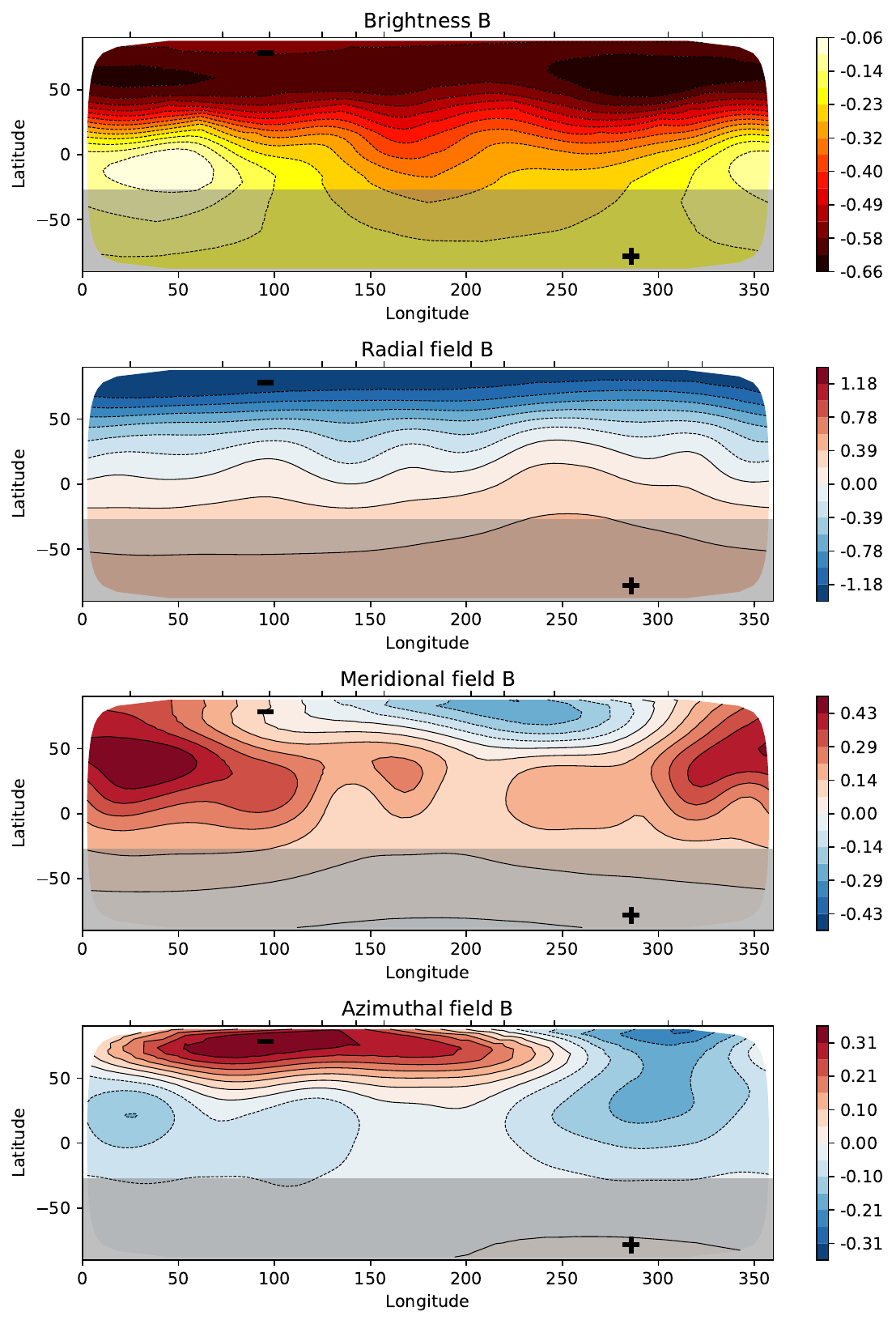}
    \caption{DQ Tau's ZDI maps of the A \textit{(left)} and B components \textit{(right)}. The top panels are showing the brightness maps resulting from the LSD Stokes I profiles fit, with a colour code representing the logarithm of the normalized brightness. The three bottom panels are showing the radial, meridional, and azimuthal magnetic field maps fitting the LSD Stokes V profiles with a colour bar indicating the field strength in kG. The latitude is going from 90$^\circ$ to $-$90$^\circ$ representing the North and South rotational poles, respectively. The grey-shaded areas represent the unseen region of the stellar surface due to the stellar inclination. The "$+$" and "$-$" signs are marking the dipolar positive and negative poles, respectively. The black ticks at the top of each map are indicating the observed longitude.}
    \label{fig:zdimaps}
\end{figure*}

The main uncertain parameter was the rotation period of the secondary (see Paper I for details).
Here again, we ran the ZDI on a grid of periods for the two components and used the reduced $\chi^2$ of Stokes V fit to constrain these periods with the 1-$\sigma$ divergence of the $\chi^2$ for the uncertainties. 
The period recovered for the primary ($P_{\rm rot}$(A) = 3.03$^{+0.03}_{-0.07}$ d) is consistent with the value derived for the sinusoidal modulation of the Kepler K2 light curve \citep{Kospal18} and with the result of the same procedure performed in Paper I.
However, the rotation period of the secondary is deviating from our previous analysis, $P_{\rm rot}$(B) = 4.53$^{+0.14}_{-0.03}$ d, but is now consistent with the periodicity observed in emission lines for our present and previous work ($\sim$4.8 d).
The reason why the convergence is optimal on different secondary's periods is not clear and may need further investigations which are outside the scope of this study.
These rotation periods mean the following observed rotational phases: 0.011, 0.331, 0.004, 0.326, 0.613, 0.253, 0.542, 0.863, 0.217, 0.549, 0.851 for the primary, and 0.683, 0.897, 0.347, 0.562, 0.436, 0.202, 0.396, 0.611, 0.847, 0.069, 0.271 for the secondary.
The resulting maps are shown ins Fig.~\ref{fig:zdimaps} and the magnetic topology is summarized in Table~\ref{tab:magTop} for the two components.
The fit are available in Appendix~\ref{ap:fitZDI}.

\begin{table}
    \centering
    \caption{Large-scale magnetic topology of DQ Tau from the ZDI analysis. The distribution is given a fraction of the total magnetic energy.}
    \begin{tabular}{l l l}
    \hline
        Distribution of & A  & B \\
        the magnetic energy & (\%) & (\%) \\
        \hline
        $\ell$=1 & 70.9 & 80.0 \\
        $\ell$=2 & 16.7 & 13.3 \\
        $\ell$=3 & 5.6 & 3.9 \\
        $\ell$=4 & 3.0 & 1.4 \\
        $\ell$=5 & 1.9 & 0.8 \\
        $\ell$=6 & 0.9 & 0.4 \\
        $\ell$=7 & 0.5 & 0.2 \\
        $\ell$=8 & 0.3 & 0.1 \\
        $\ell$=9 & 0.1 & 0.0 \\
        $\ell$=10 & 0.0 & 0.0 \\
        poloidal field & 41.6 & 94.9 \\
        toroidal field & 58.4 & 5.1 \\
        axisymmetry (|$m$|<$\ell$/2) & 31.0 & 87.6 \\
        \hline
        Magnetic field strengths & (kG) & (kG) \\
        \hline
        $\langle B \rangle_V$ & 0.210 & 0.396 \\
        $B_{\rm max}$ & 0.605 & 1.458 \\
        $B_{\rm dip,max}$ & 0.338 & 0.672 \\

    \hline
    \end{tabular}
    \label{tab:magTop}
\end{table}

The resulting brightness maps are similar to the ones obtained in Paper I, showing a large dark feature around 60$^\circ$ latitude for the two components.
The magnetic topology of the B component is very close to our previous work as well, largely dominated by the poloidal field and by the dipolar components.
The magnetic field strengths are similar, slightly higher for the mean strength, and the dipole negative pole is located at 78$^\circ$ latitude and 95$^\circ$ longitude.

The A component magnetic topology seems to have changed significantly in the contribution of the toroidal field, which is now dominating the magnetic energy (from 33 to 58\% of the total magnetic energy).
The dipolar component contribution remains similar and the magnetic field strengths slightly higher than in our previous work. 
The dipole negative pole is now located at a latitude of 54$^\circ$ and a longitude of 28$^\circ$.

\subsubsection{Zeeman intensification}
\label{sec:small-scale}

We investigate the properties of the small-scale magnetic fields on the components of DQ Tau. This is done following the same approach that was used on DQ Tau and other T Tauri binaries \cite[][Paper I]{Hahlin22}. In order to obtain separate spectra for each component, we use a spectral disentangling procedure described in \cite{Folsom10} with additional capacity to simultaneously disentangle the telluric signal \citep[see][]{Kochukhov19}. This was done for the wavelengths between 963 and 982\,nm, this wavelength range contains a particularly useful set of \ion{Ti}{I} lines that originate from the same multiplet and have lines with a large variation in magnetic sensitivity. Due to the radial velocity sampling, a few of the \ion{Ti}{I} lines ended up blended with tellurics making their profiles distorted, for this reason, they were not included in the inference.

We generated a grid of synthetic spectra using the polarised radiative transfer code \texttt{SYNMAST} \citep{Kochukhov10}, line lists from VALD, and \texttt{MARCS} model atmospheres \citep{Gustafsson08}. We used the SoBAT library \citep{Anfinogentov21} for IDL to do MCMC sampling in order to find the small-scale magnetic field assuming a multi-component model. This is done by combining synthetic spectra $S_{i}$ of different magnetic field strengths based on the fraction of the stellar surface covered by each field strength,
\begin{equation}
    S=\sum_{i}f_{i}S_{i}.
\end{equation}
Using the same principle, the average magnetic field $\langle B\rangle_{I}$ can be determined from,
\begin{equation}
    \label{eq:Bavg}
    \langle B\rangle_{I}=\sum_{i}f_{i}B_{i}.
\end{equation}
In this model, the filling factors $f_{i}$ are the free magnetic parameters. We fitted the disentangled spectra of both components simultaneously using non-magnetic stellar parameters of titanium abundance, luminosity ratio (LR), rotational velocity, and radial velocity. The latter two properties are separate for each component while the former are shared parameters.

In principle it is possible to add an arbitrary number of filling factors, this does however cause an issue as a small contribution of strong magnetic fields can have a significant impact on the average field strength on the stellar surface without significantly impacting the quality of fit \citep[e.g.][]{Shulyak19}. In order to avoid this we used the Bayesian information criterion \citep[BIC,][]{Sharma17} to only include filling factors that significantly improved the fit. This was done by iteratively adding filling factors corresponding to increasing magnetic field strength until the BIC no longer increased in value. For DQ Tau we found that a multi-component model with 4 components corresponding to 0, 2, 4, and 6\,kG is the most suitable description.

\begin{figure*}
    \centering
    \includegraphics[width=\textwidth]{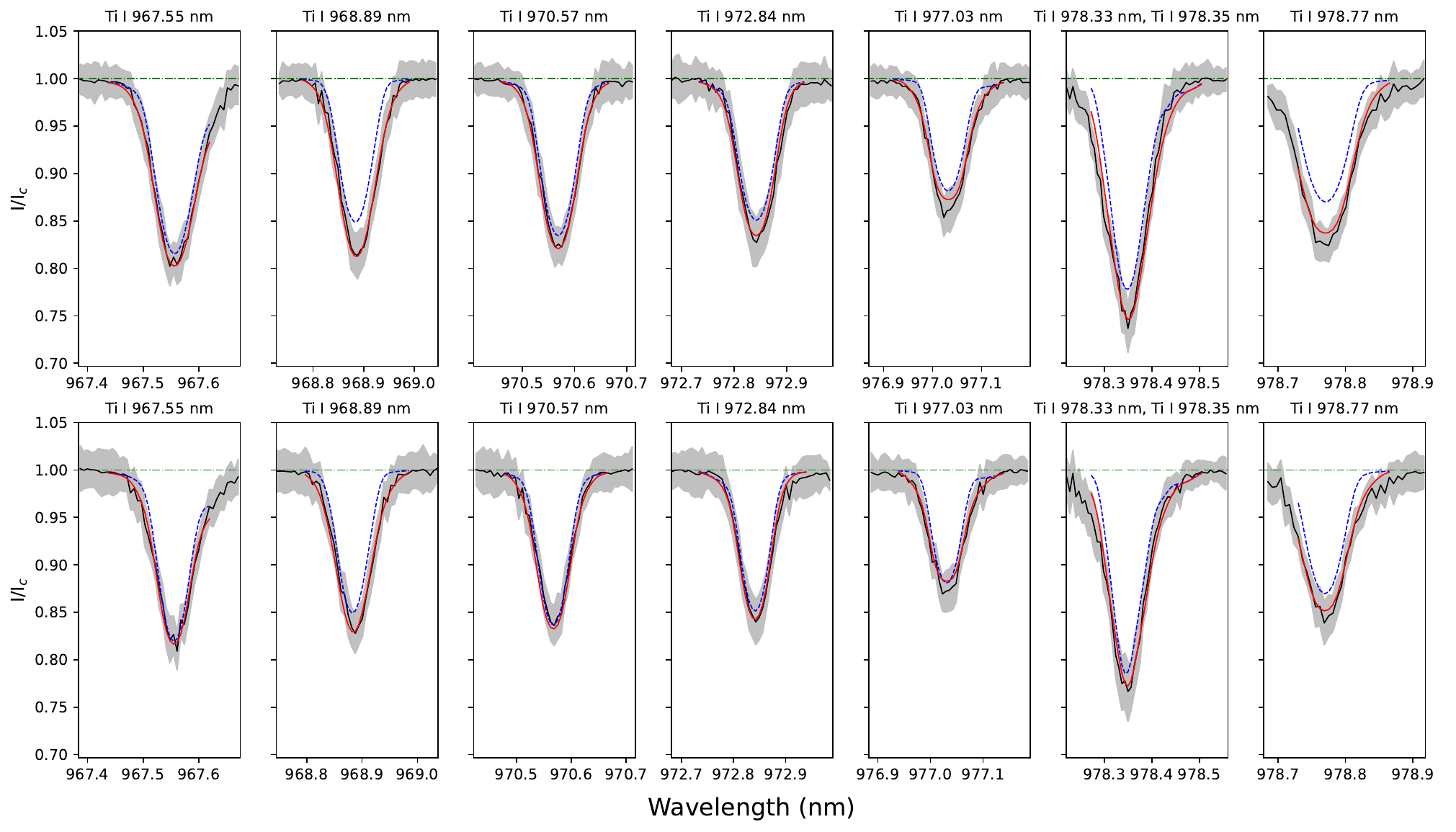}
    \caption{Fit to the disentangled spectra of DQ Tau. Solid red line shows the best fit to the observations in black while the blue dashed line shows the non-magnetic spectra with otherwise identical stellar parameters. Shaded area marks the uncertainties obtained from the observations. The top row is the disentangled spectrum of component A while the bottom is for component B.}
    \label{fig:DQTauTilines}
\end{figure*}

\begin{figure}
    \centering
    \includegraphics[width=\linewidth]{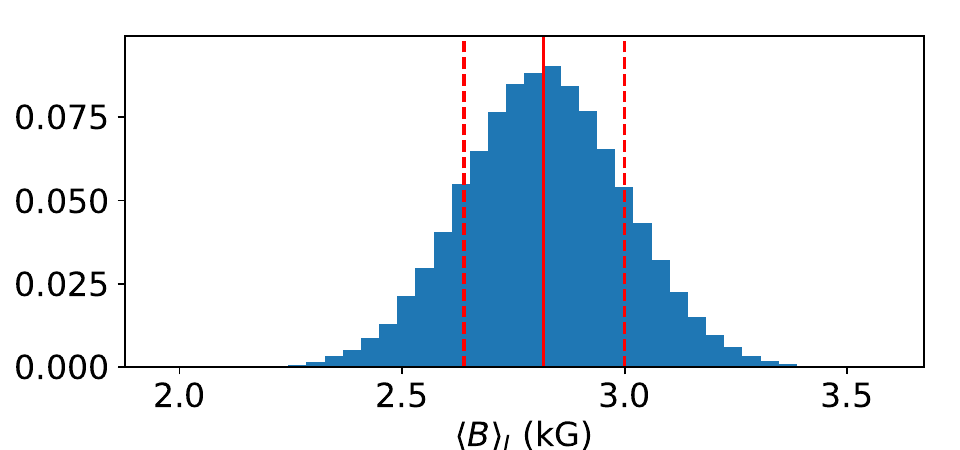}
    \includegraphics[width=\linewidth]{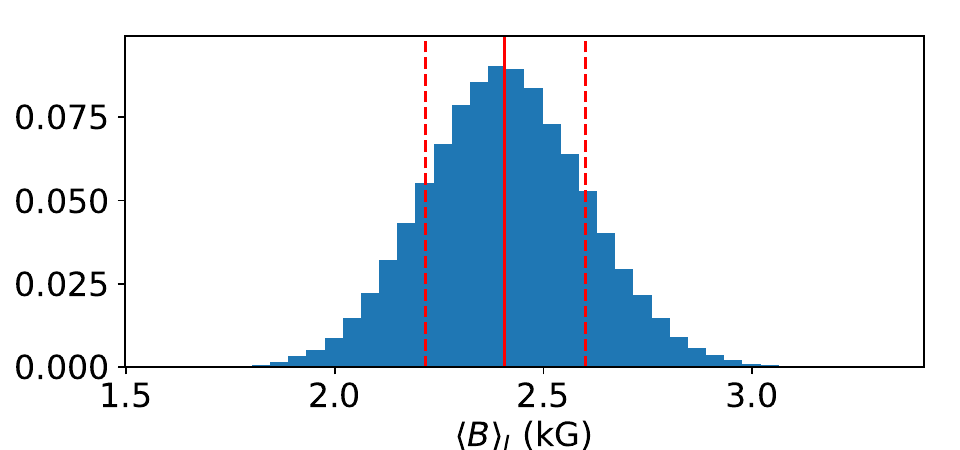}
    \caption{Distribution of the average magnetic field for the components of DQ Tau. Solid line represents the median and the dashed lines represents 68\,\% credence regions. The top and bottom is for the A and B component respectively.}
    \label{fig:DQTauBdist}
\end{figure}

We run the code until it reached an effective sample size \citep[ESS,][]{Sharma17} of 1000. The resulting fit can be seen in Fig. \ref{fig:DQTauTilines} and the corresponding inference parameters are in Table \ref{tab:small-scale} and Fig \ref{fig:DQTauSmallScale}. The obtained magnetic field strengths of the components are DQ Tau are $2.82\pm0.18$ and $2.41\pm0.19$\,kG, their distributions can be seen in Fig.~\ref{fig:DQTauBdist}. The values are similar to previous values reported in \cite{Pouilly23}, the B component is essentially unchanged while we see a small, but statistically significant, increase in the field strength of the A component. We do note a lower abundance compared to the previous result, this is likely due to the increased accretion reported in Sect. \ref{subsubsec:EmLineDQ}, which would reduce line depths due to veiling. 

\begin{table}
    \centering
    \caption{Small-scale magnetic field parameters obtained for DQ Tau. Note that $\langle B\rangle$ is the median of Eq. \ref{eq:Bavg} calculated for each step, it may vary slightly from the median value filling factors reported here. For the complete set of non-magnetic parameters, see Fig.~\ref{fig:DQTauSmallScale}.}
    \begin{tabular}{l ll}
    \hline
    Parameter & A & B \\
    \hline
    $\langle B\rangle_I$ (kG) & $2.82\pm0.18$ & $2.41\pm0.19$ \\
    $f_{2}$ & $0.44\pm0.11$ & $0.31\pm0.10$ \\
    $f_{4}$ & $0.07\pm0.09$ & $0.03\pm0.05$\\
    $f_{6}$ & $0.27\pm0.05$ & $0.27\pm0.04$\\ 
    \hline
    \end{tabular}
    
    \label{tab:small-scale}
\end{table}

\subsection{AK Sco }

Here we present the analysis of the AK Sco's dataset.
This is done in a similar way to DQ Tau, in order to compare the behaviour of the two systems.

\subsubsection{LSD profiles and radial velocity}
\label{subsubsec:LSDAK}

\begin{figure*}
    \centering
    \includegraphics[width=.49\textwidth]{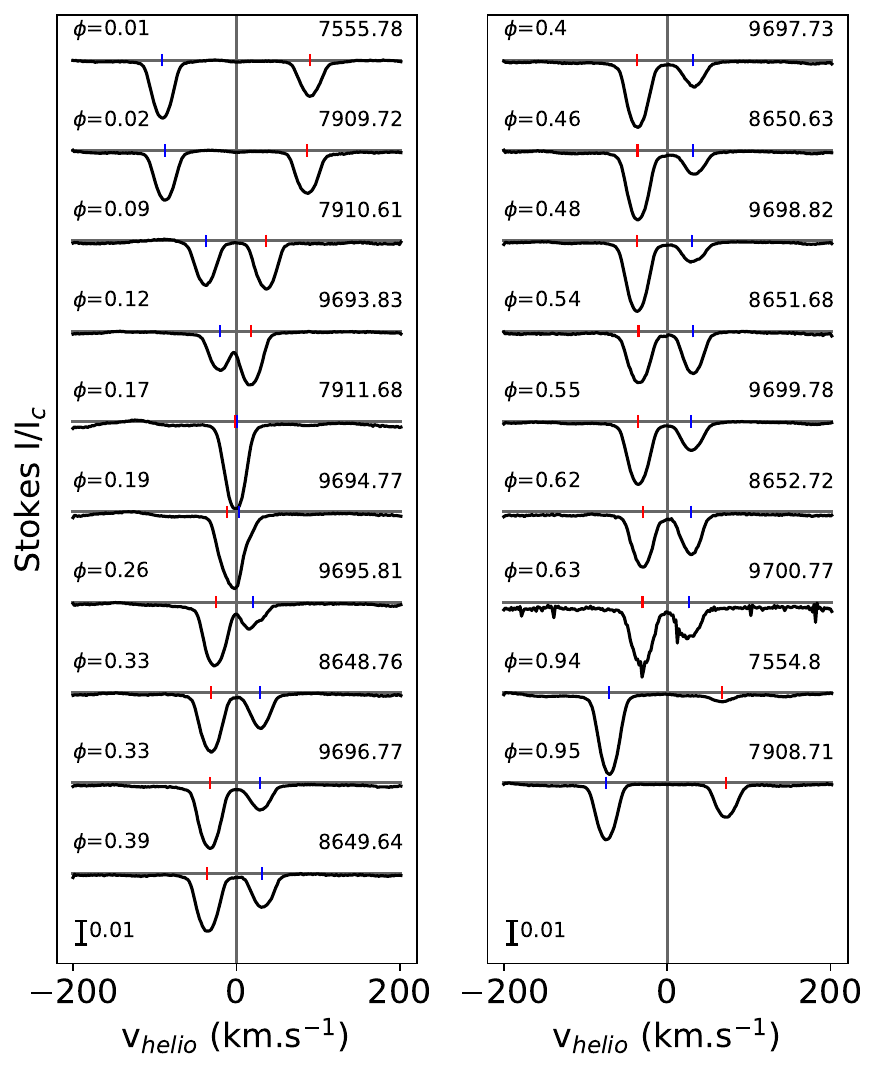}
    \includegraphics[width=.49\textwidth]{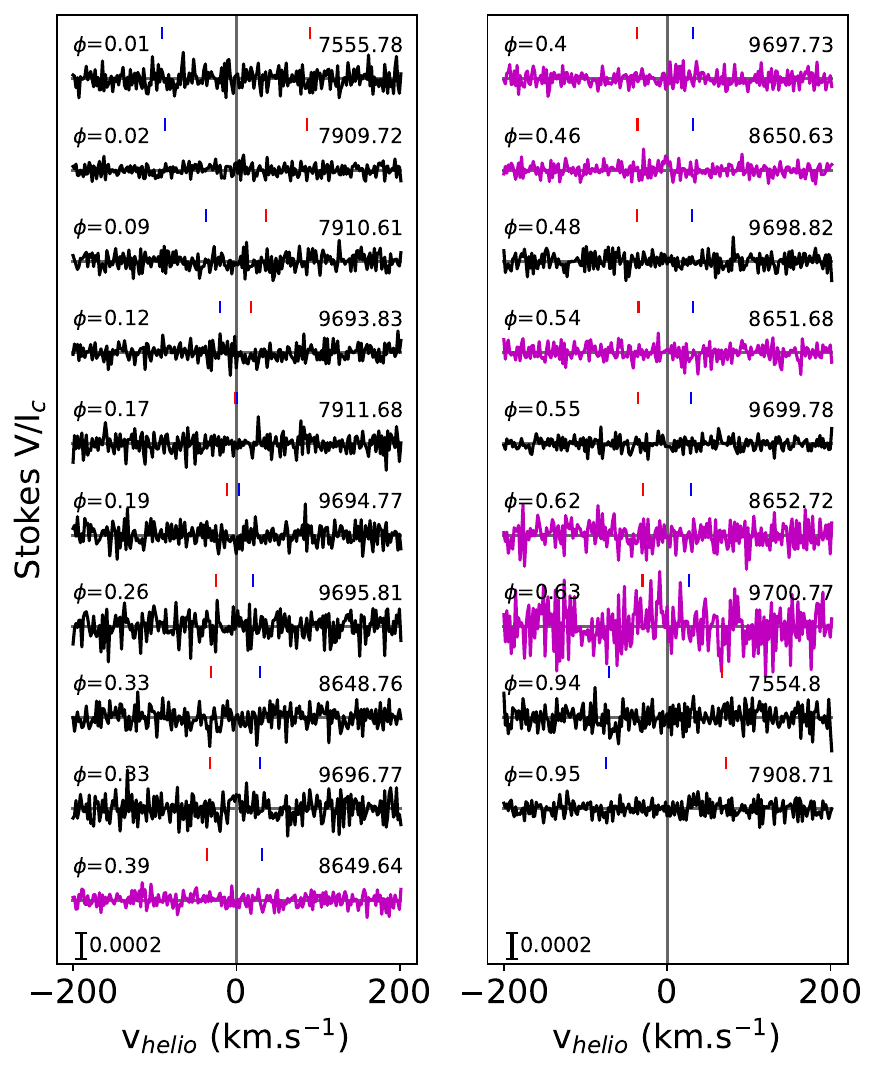}
    \caption{Same as Fig.~\ref{fig:lsdprof} for AK Sco. The magenta Stokes V profiles have been divided by 2 for more visibility.}
    \label{fig:lsdprofAK}
\end{figure*}

The LSD profiles of AK Sco were computed using the same procedure as DQ Tau (Sect.\ref{subsubsec:LSD} and Paper I) using a line list adapted to the higher effective temperature, except that only a few observations were acquired with 4 sub-exposures. 
The others have only 2 sub-exposures available, meaning that we cannot compute the Null profiles, that cancel the polarisation signal for these observations. 
The FAP computed from the prescription of \cite{Donati92} clearly indicate no definite Stokes V signal detection in any of the observations (FAP = 1 for all observations).
The LSD Stokes I and V profiles are displayed in Fig~\ref{fig:lsdprofAK}, and exhibit a S/N between 1836 and 2807 for the Stokes I, and between 2469 and 19241 for the Stokes V profiles, computed from about 9000 photospheric lines.

\begin{figure}
    \centering
    \includegraphics[width=.45\textwidth]{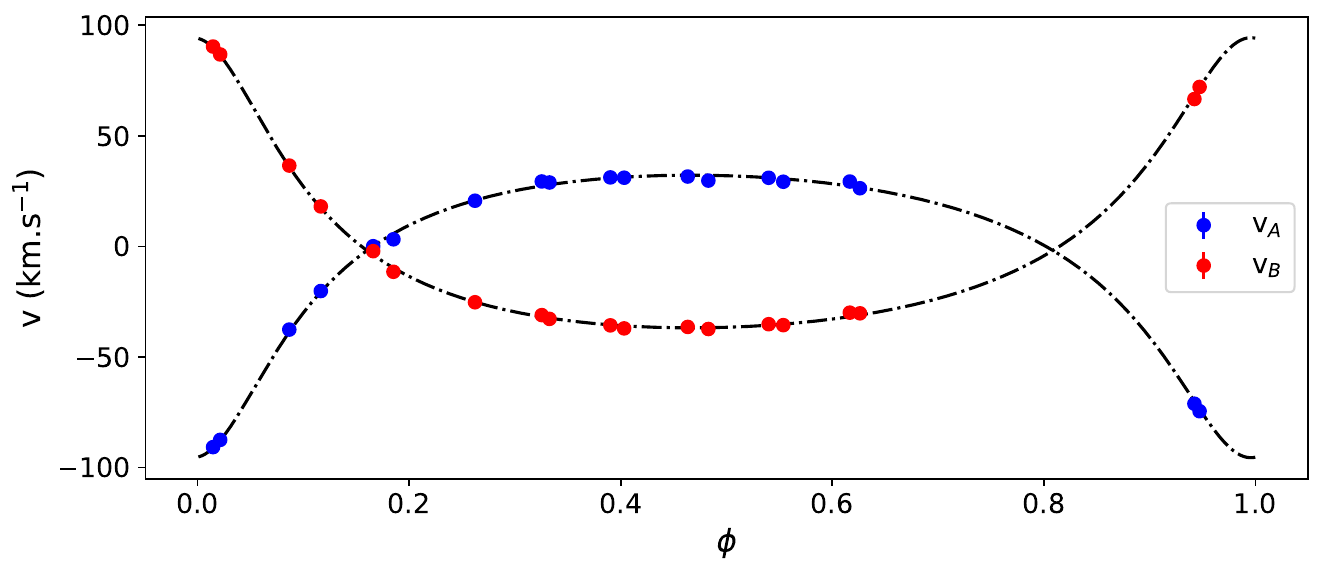}
    \caption{Same as Fig.~\ref{fig:vrad} for the AK Sco system.}
    \label{fig:vradAK}
\end{figure}

\cite{Jarvinen18} noticed that the secondary component was very faint at HJD 2~457~554.8 and interpreted it as a dust occultation by a small-scale cloud. 
We do not observe such an occultation at the same orbital phase one year later (HJD 2~457~908.71).
Furthermore, we notice the inversion of the equivalent width ratio between the two components around $\phi$ = 0.12 and $\phi$ = 0.94, observed by \cite{Alencar03} as well, and here again, interpreted by the presence of dust resulting in a higher extinction on the component behind the other one in the line of sight.
Then we derived the radial velocity of each component and the orbital elements of the system from the same procedure as in Sect.~\ref{subsubsec:LSD}, meaning a disentangling routine of the LSD Stokes I profiles using the orbital solution of \cite{Alencar03} as a first guess, and an LMA fit of the obtained radial velocities.
The results are summarised in Table~\ref{tab:vradAK} and \ref{tab:orbElementsAK}, and displayed in Fig.~\ref{fig:vradAK}.

\begin{table}
    \centering
        \caption{Same as Table~\ref{tab:vrad} for AK Sco's components}
    \begin{tabular}{l | l l | l l}
        \hline
        HJD & \vrad(A) & $\delta$\vrad(A) & \vrad(B) & $\delta$\vrad(B) \\
        ($-$2~450~000 d) & \kms & \kms & \kms & \kms \\
        \hline
        
        7554.79823 & -71.16 & 1.09 & 66.65 & 1.41 \\
        7555.78259 & -90.78 & 0.31 & 90.39 & 0.36 \\
        7908.70690 & -74.55 & 0.29 & 72.11 & 0.41 \\
        7909.71661 & -87.50 & 0.14 & 86.81 & 0.17 \\
        7910.60562 & -37.63 & 0.28 & 36.54 & 0.27 \\
        7911.68494 & 0.08 & 0.40 & -2.16 & 0.35 \\
        8648.75645 & 29.36 & 0.51 & -31.08 & 0.48 \\
        8649.63905 & 31.20 & 0.51 & -35.72 & 0.46 \\
        8650.63415 & 31.59 & 1.02 & -36.42 & 1.06 \\
        8651.67538 & 31.02 & 0.21 & -35.19 & 0.22 \\
        8652.71917 & 29.33 & 0.29 & -29.98 & 0.23 \\
        9693.83067 & -20.19 & 0.45 & 18.07 & 0.45 \\
        9694.76515 & 3.20 & 0.41 & -11.48 & 0.41 \\
        9695.81443 & 20.66 & 0.94 & -25.26 & 0.99 \\
        9696.77257 & 28.90 & 0.86 & -32.79 & 0.88 \\
        9697.73313 & 31.06 & 0.92 & -37.09 & 0.93 \\
        9698.81650 & 29.77 & 1.11 & -37.36 & 1.15 \\
        9699.78042 & 29.24 & 0.73 & -35.62 & 0.72 \\
        9700.76676 & 26.30 & 0.78 & -30.27 & 0.73 \\

    \hline
    \end{tabular}
    \label{tab:vradAK}
\end{table}

\begin{table}
    \centering
    \caption{Same as Table~\ref{tab:orbElements} for the AK Sco system.}
    \begin{tabular}{l c c}
        \hline
        Parameter & This work & \cite{Alencar03} \\
        \hline
        $P_{\rm orb}$ (days)& 13.6093 $\pm$ 0.0001 & 13.609453 $\pm$ 0.000026 \\
        $\gamma$ (\kms) & -1.86 $\pm$ 0.19 & -1.97 $\pm$ 0.10 \\
        $K_1$ (\kms) & 63.89 $\pm$ 0.46 & 64.45 $\pm$ 0.23 \\
        $K_2$ (\kms) & 65.61 $\pm$ 0.46 & 65.32 $\pm$ 0.24 \\
        $e$ & 0.469 $\pm$ 0.004 & 0.4712 $\pm$ 0.0020 \\
        $\omega$ ($^{\circ}$) & 184.98 $\pm$ 0.86 & 185.40 $\pm$ 0.33 \\
        $T_{\rm peri}$ (HJD - 2~400~000) & 57552.73941 $\pm$ 0.1502 & 46654.3634 $\pm$ 0.0086 \\
    \hline
    \end{tabular}
    \label{tab:orbElementsAK}
\end{table}

\noindent The orbital parameters obtained are perfectly consistent with the work of \cite{Alencar03}, indicating that the system's motion seems very stable, which confirms that the usage of a data set unevenly sampled over 6 years is valid for such analysis.
In particular, the argument at periastron obtained by \cite{Alencar03} from the data obtained between 1986 and 1994 and the one fitted here from the data obtained between 2016 and 2022 is consistent within the uncertainties.
Therefore, as opposed to DQ Tau, AK Sco does not show any apsidal motion.

\subsubsection{TESS light curves}

\begin{figure*}
    \centering
    \includegraphics[width=.48\textwidth]{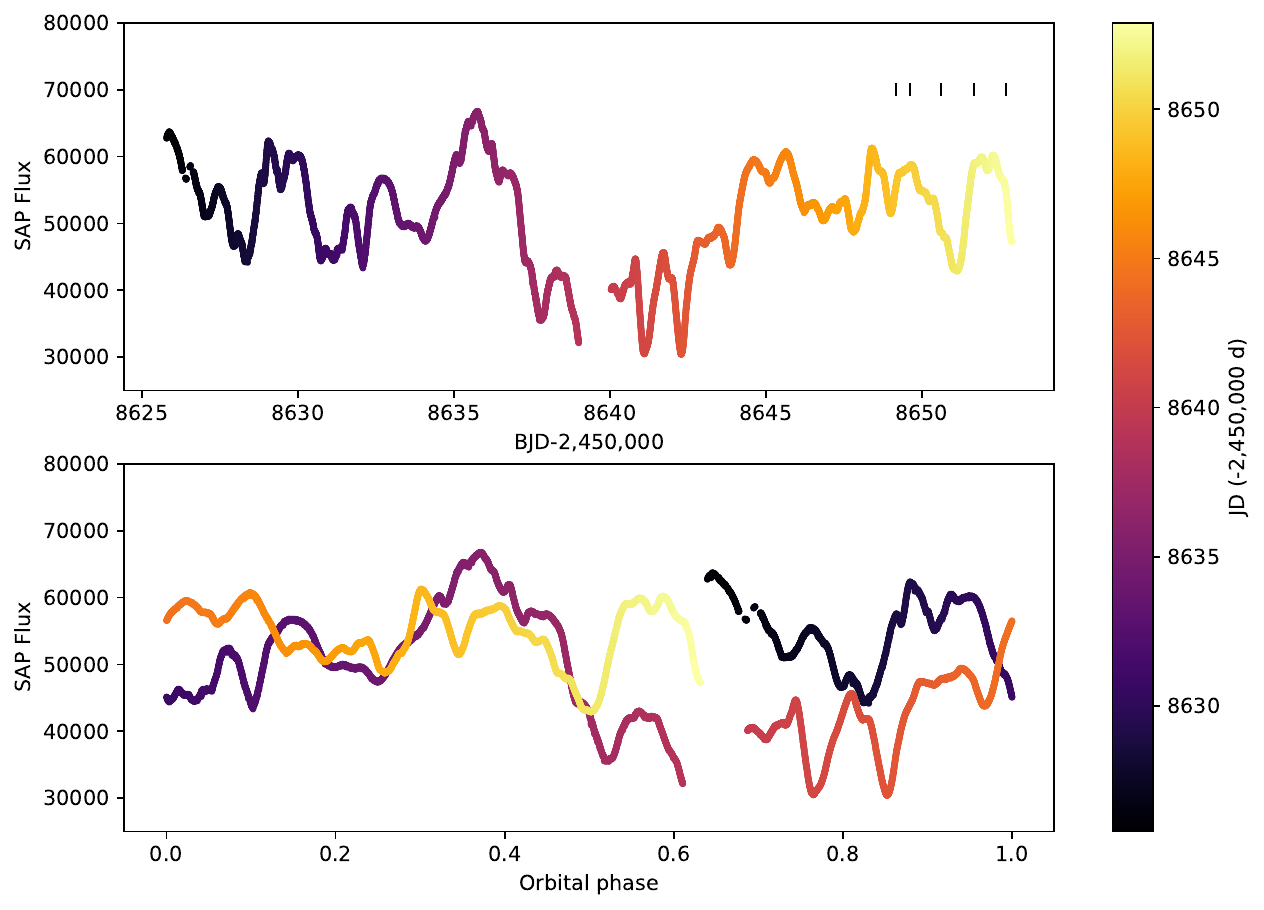}
    \includegraphics[width=.48\textwidth]{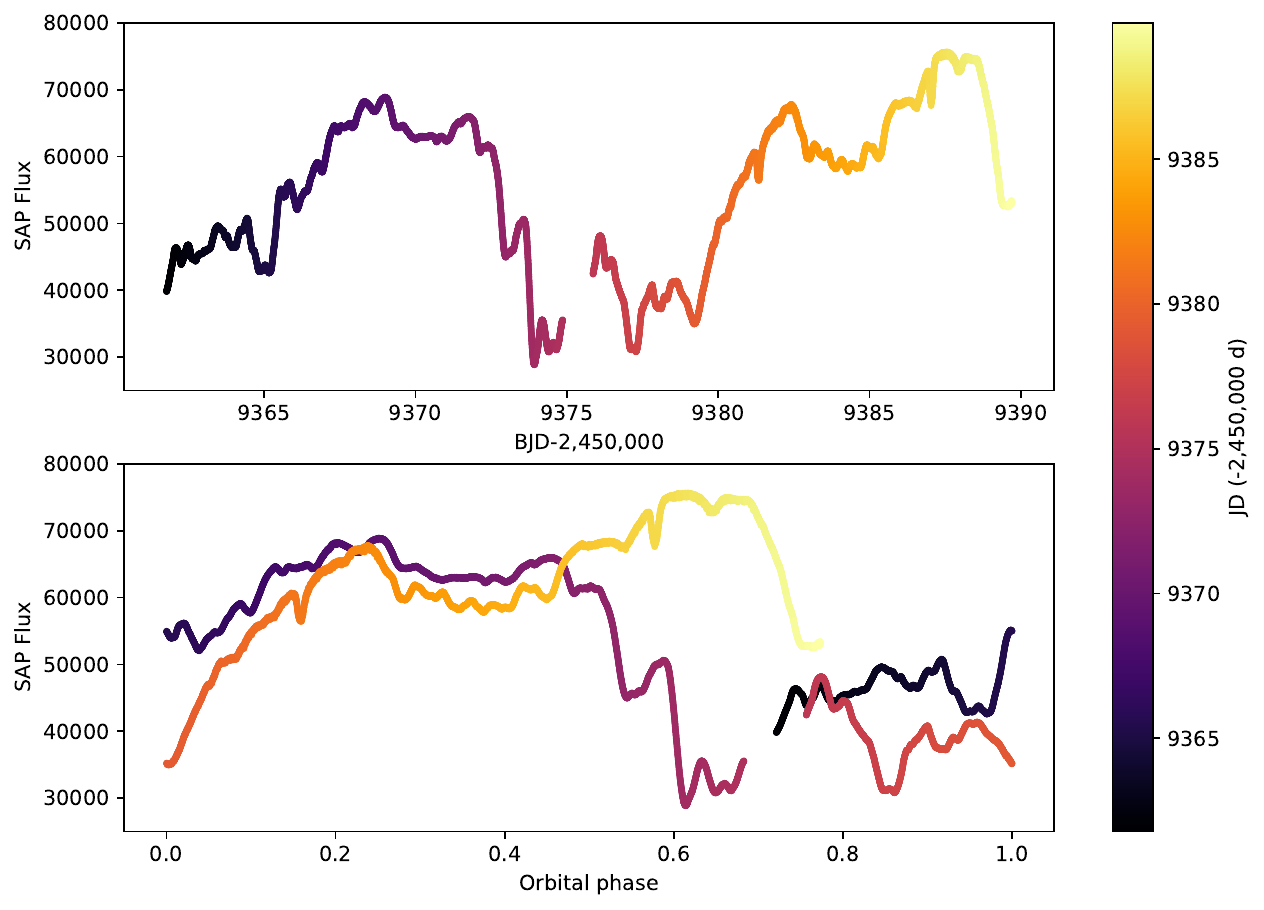}

    \caption{TESS light curves of AK Sco acquired in 2019 \textit{(left)} and 2021 \textit{(right)}. The top panels are displayed as a function of the BJD and the bottom panels are folded with the orbital period of 13.6 d. The colours are scaling the BJD. The black ticks on the 2019 light curve indicate the simultaneous HARPSpol observations.}
    \label{fig:tess}
\end{figure*}

TESS observed AK Sco during 2 periods in 2019 and 2021, each time for 27 days.
The resulting light curves are shown in Fig.~\ref{fig:tess} and display very different behaviours.
The 2019 observations exhibit at quasi sinusoidal modulation on an $\sim$3-day period according to the Lomb-Scargle periodogram analysis, with a dimming event occurring between HJD 2~458~637 and HJD 2~458~644. 
The 2021 observations consist of a lower frequency and higher amplitude modulation, similar to the dimming event of 2019.
On both light curves folded with the 13.6 d orbital phase, one can notice that the minimum is occurring on one orbital cycle only right after the apastron until the periastron (from phase $\sim$0.5 to 1.0).
The second half of the 2021 light curve is showing a similar drop, but a bit later, between phase 0.7 and 1.0.
Part of the HARPSpol observations was acquired simultaneously with TESS and seems to occur during a rather quiet period, meaning outside the largest minima and maxima observed on other cycles.

\subsubsection{H$\alpha$ line}

\begin{figure}
    \centering
    \includegraphics[width=.46\textwidth]{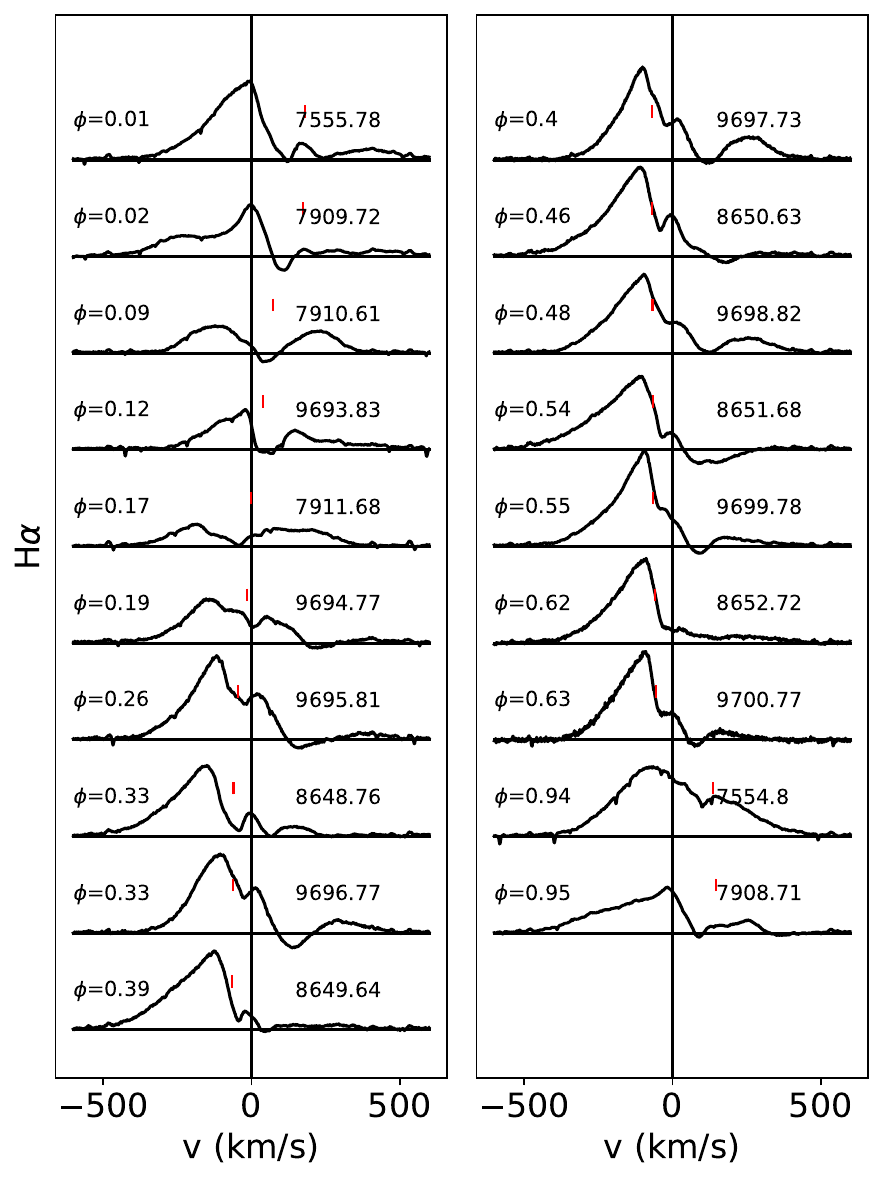}
    \caption{Same as Fig.~\ref{fig:ha} for AK Sco's H$\alpha$ lines.}
    \label{fig:haAK}
\end{figure}

The photospheric template used to compute the residual spectrum is Par-1646, studied by \cite{Villebrun19} and showing a \teff\ = 6310 K and a \vsini\ = 15.67 \kms.
The only Balmer line in emission in the residual spectrum is H$\alpha$, which might indicate a lower accretion rate in AK Sco than in DQ Tau, we thus focused our analysis on this line.
The resulting profiles are shown in Fig.~\ref{fig:haAK}, corrected for the primary's radial velocity. 
They exhibit a large variation along the orbital cycle, but from cycle to cycle as well, as shown by Fig.~\ref{fig:haAKPeri}.

\noindent Most of the profiles show two main peaks, separated by a central absorption which is always located at the heliocentric velocity.
A redshifted absorption can be seen at $\phi$ = 0.19, 0.26, 0.33, 0.46, 0.54, 0.55, and 0.63, from about 180 to 300, 120 to 250, 80 to 200, 150 to 250, 50 to 300, 50 to 120, and from 50 to 100 \kms, respectively.
Such a feature, usually ascribed to material falling on the stellar surface, is called Inverse P Cygni (IPC) profile and is commonly observed on cTTSs \citep[e.g.,][]{Pouilly20, Pouilly21}.
A similar pattern can be observed at a similar velocity at $\phi$ = 0.39 but does not go below the continuum level.
At $\phi$ = 0.02, the blue wing is strongly weakened (~-150 \kms), in particular when compared to the other profile at $\phi$ = 0.01 a few cycles earlier (Fig.~\ref{fig:haAKPeri}).
This signature is reminiscent of the inter-cycle variation of the accretion-ejection processes reported in the literature \citep{Alencar03, GomezDeCastro20}.

The 2022 data set was acquired during 8 consecutive nights, making these observations suitable for more detailed variability analysis.
We performed a 2D-periodogram analysis of these H$\alpha$ lines, yielding the detection of two periodicities on two different regions of the line.
The 13-day orbital period is recovered at $\sim$ -100 \kms, with a FAP of about 10$^{-2}$, and a second signal is detected (FAP$\sim$10$^{-3}$) around 200 \kms (i.e., the region of the IPC profiles), at half of the orbital period.
As the stars are expected to be tidally synchronised \citep{Alencar03}, recovering half of the orbital period in a region where the accretion is dominating the line's variability means that we detect the accretion of both components during this part of the orbital cycle.

\begin{figure}
    \centering
    \includegraphics[width=.45\textwidth]{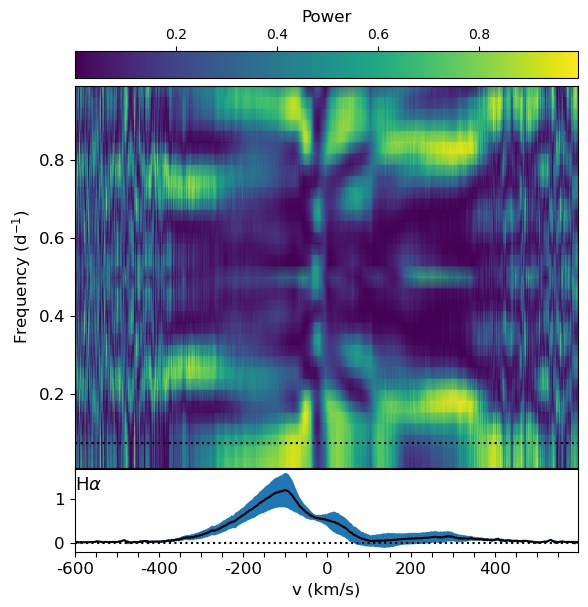}
    \caption{\textit{Top :} 2D periodogram of the 2022 H$\alpha$ lines of AK Sco, centred on the primary's velocity. The x- and y-axis represent the line velocity and the frequency of the periodogram. The colours are scaling the power of the periodogram and the horizontal dotted line is showing the orbital period of the system. The signals at frequencies higher than 0.5 d$^{-1}$ are aliases due to the $\sim$1-day sampling of the time series. \textit{Bottom :} Mean profile \textit{(black)} and its corresponding variance \textit{(blue)}. The x-and y-axis represent the line velocity and the normalised flux of the line.}
    \label{fig:HaP2D}
\end{figure}

\begin{figure}
    \centering
    \includegraphics[width=.45\textwidth]{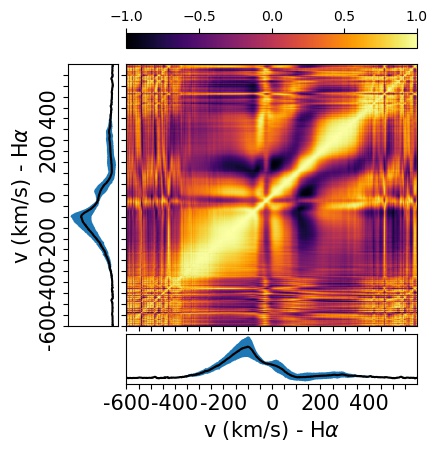}
    \caption{Same as Fig.~\ref{fig:CMhe} for 2022 AK Sco's residual H$\alpha$ line.}
    \label{fig:HaCM}
\end{figure}

The H$\alpha$ autocorrelation matrix, shown in Fig.~\ref{fig:HaCM}, revealed four main correlated substructures in the line: from -300 to -70, 0 to 100, 120 to 180, and from 200 to 300 \kms. 
The two latter are consistent with the redshifted absorptions observed in the profile.
The fact these substructures are not correlated between them suggests that two different physical processes are governing the variability of these two regions.
This was expected from the half-orbital period modulation of this region, suggesting that both components show this accretion signature.
However, we can notice that the 120 to 180 \kms\ substructure is anti-correlated with the blueshifted part of the line.
Given the overall blueshift of the mean profile, one can guess that H$\alpha$ is mainly emitted by the secondary during these phases, added to the emission excess in the blue wing of the line produced simultaneously with the redshifted absorption by the infalling material.
This suggests that this particular substructure can be ascribed to the secondary's IPC profile when the one between 200 and 300 \kms\ is produced by the primary.

\begin{figure}
    \centering
    \includegraphics[width=.46\textwidth]{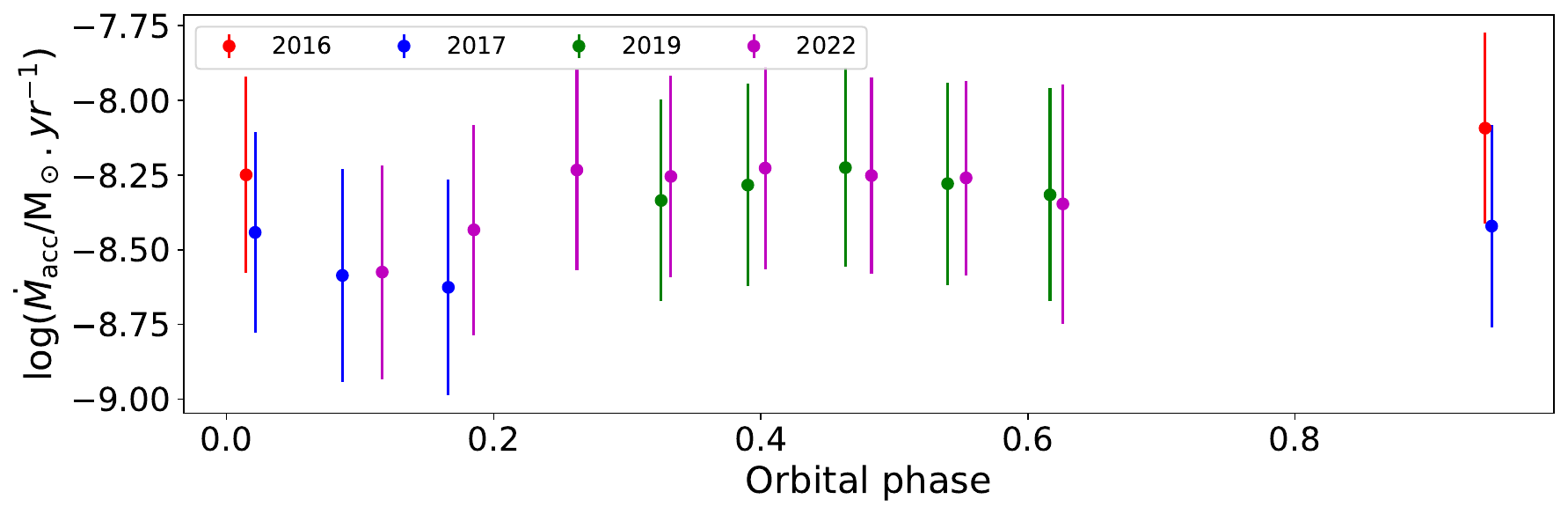}
    \caption{Mass accretion rate of AK Sco computed from the H$\alpha$ lines. The red, blue, green, and magenta markers are indicating the 2016, 2017, 2019, and 2022 observations, respectively.}
    \label{fig:MacchaAK}
\end{figure}

Finally, we computed the mass accretion rate of the system using the same methodology as in Sect.~\ref{subsubsec:EmLineDQ}, they are displayed in Fig.~\ref{fig:MacchaAK}.
The mass accretion seems modulated on the orbital period, showing maxima at the periastron and apastron. 
However, due to the IPC profiles reported between phase 0.2 and 0.6, the corresponding mass accretion rate values, which depend on the line's equivalent width, might be underestimated.
The mean mass accretion thus obtained is about 10$^{-8.3}$ \msunyr, from a mean H$\alpha$'s EW of 6.34 $\pm$ 0.09 \AA\ which is consistent with the study of \cite{Alencar03}.

\subsubsection{Zeeman intensification}
\label{subsubsec:ZintensAK}

We apply the same method to obtain small-scale magnetic fields on AK Sco that was used on DQ Tau and described in Sect. \ref{sec:small-scale}. The difference is that we use HARPSpol spectra, which do not reach sufficiently high wavelengths to observe the \ion{Ti}{I} multiplet used for DQ Tau. Instead, we use lines from a multiplet of \ion{Fe}{I} lines identified by \cite{Kochukhov20} as suitable lines for magnetic investigation. These lines are located at around 550\,nm and contain both magnetically sensitive and insensitive lines. 

An issue with the observations of AK Sco can be seen in the Stokes I LSD profiles of Fig. \ref{fig:lsdprofAK}, it appears as if the relative strength between the two components varies significantly between different observations. This was attributed by \cite{Alencar03} to obscuration due to surrounding dust that causes the more distant component to be more obscured. This presents an issue for our spectral disentangling as the key assumption is an unchanging spectrum as a function of time. While dust obscuration is a continuum effect that should affect all lines in a similar way, this variation might cause some systematic errors in the disentangling process. In order to verify the robustness of our measurement, we perform spectral disentangling twice with different phase coverage to obtain two sets of spectra that can be compared. The first set is composed of spectra observed at the orbital phases below $\sim 0.2$ and above $\sim 0.8$ (referred to as Sample 1). In the second set, we add a few spectra from the intermediate orbital phases (Sample 2). We found that adding too many spectra with similar radial velocities causes issues in disentangling, we also avoid spectra with distortions or extreme deviation in line strength based on the LSD profiles (e.g. $\phi=0.94$ in Fig. \ref{fig:lsdprofAK}). 

In addition to the non-magnetic parameters described in Sect. \ref{sec:small-scale}, we also allow the continuum to shift around each line. This is because we find that, depending on the selection of phases for spectral disentangling, the continuum of the disentangled spectra is somewhat affected in regions with a large number of close lines. 

When looking at hotter stars, NLTE effects can affect the strength and shape of spectral lines. In the context of magnetic fields, \cite{Hahlin23} found the impact of NLTE effects on sun-like stars at temperatures above $\sim$6000\,K starts to become more significant. For this reason, it is worthwhile to investigate the impact of NLTE on the lines used in this study. This was done by calculating spectra with and without departure coefficients from \cite{Amarsi22} and comparing the equivalent widths of our lines of interest. We found that the \ion{Fe}{I} lines used in this study are not significantly affected by NLTE effects as the change in equivalent width is about 1\,\% for each line. Regardless of its small impact, we decided to include it as the non-detection of large-scale fields in the Stokes V profiles (see Fig. \ref{fig:lsdprofAK}) indicates weak fields. This means that even small NLTE effects could potentially change the obtained magnetic field significantly.

\begin{figure}
    \centering
    \includegraphics[width=\linewidth]{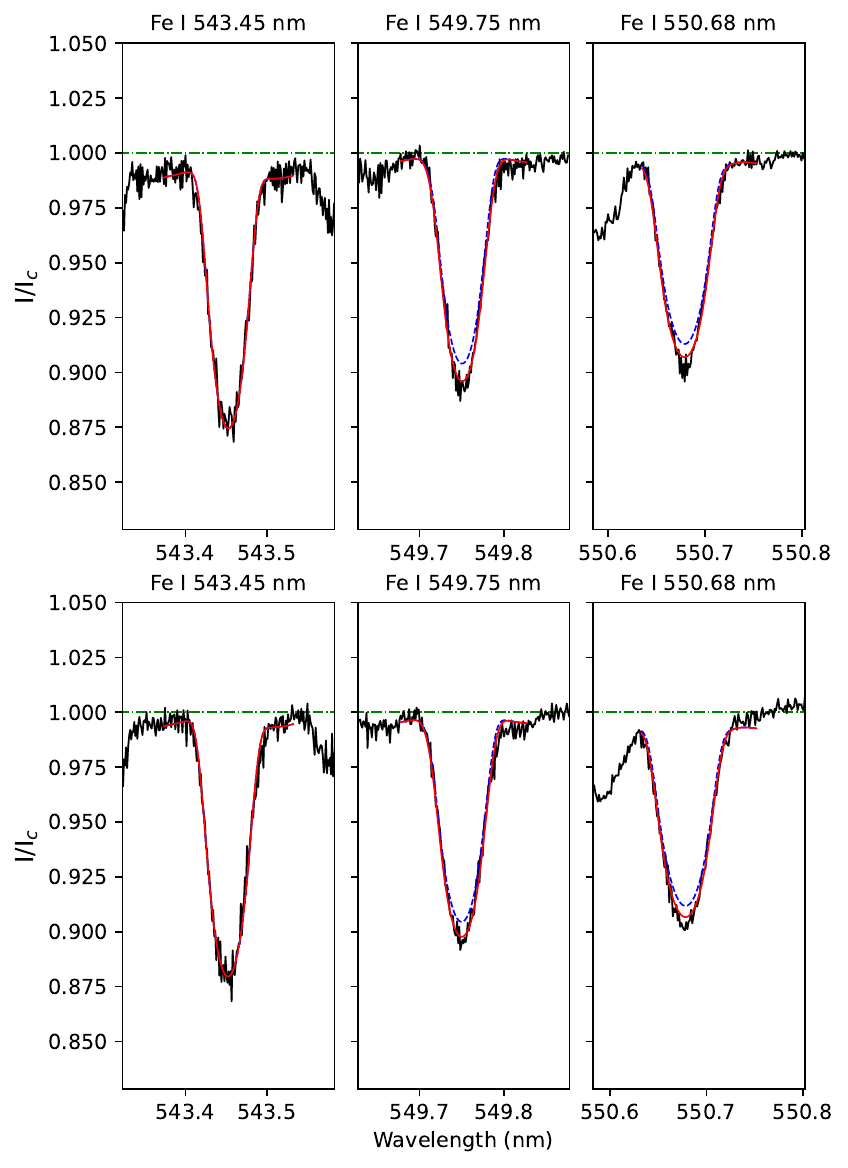}
    \caption{Fit to the disentangled spectra of Sample 2 of AK Sco. Solid red line shows the best fit to the observations in black while the blue dashed line shows the non-magnetic spectra with otherwise identical stellar parameters. The top row is the disentangled spectrum of component A while the bottom is for component B.
}
    \label{fig:AKScoTilines}
\end{figure}

We then run the inference in the same way as described in Sect. \ref{sec:small-scale}.
We found that the most suitable model describing the magnetic field is a two-component model with field strengths of 0 and 2 kG. The obtained median parameter fit can be seen in Fig. \ref{fig:AKScoTilines} with the magnetic results summarised in Table \ref{tab:small-scaleAK} and the remaining parameters given in Fig. \ref{fig:AKScoSmallScale}. We find field strengths close to 1\,kG for both components regardless of phase sampling. The A component appears unaffected while the B component changes field strength by $\sim0.2$\,kG between Sample 1 and Sample 2. While most other inference parameters are unaffected by phase sampling, the LR does change significantly. This is however expected due to the dust obscuration observed in LSD profiles.

\begin{figure}
    \centering
    \includegraphics[width=\linewidth]{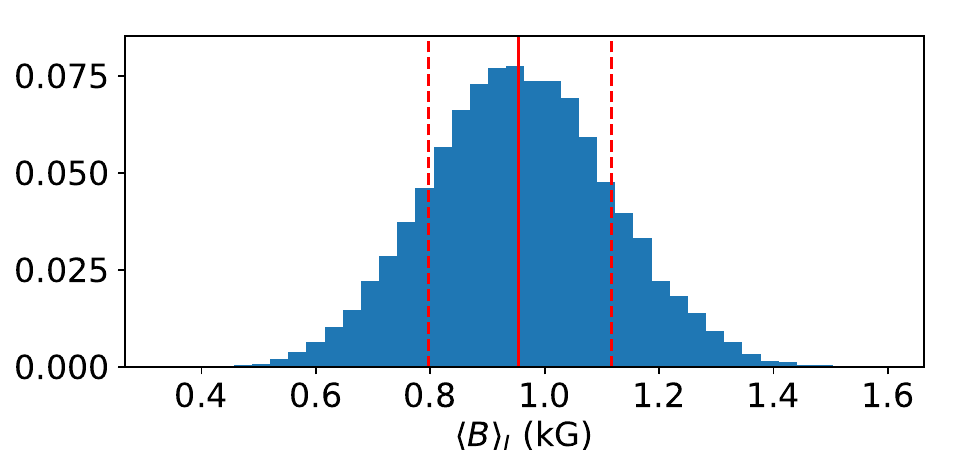}
    \includegraphics[width=\linewidth]{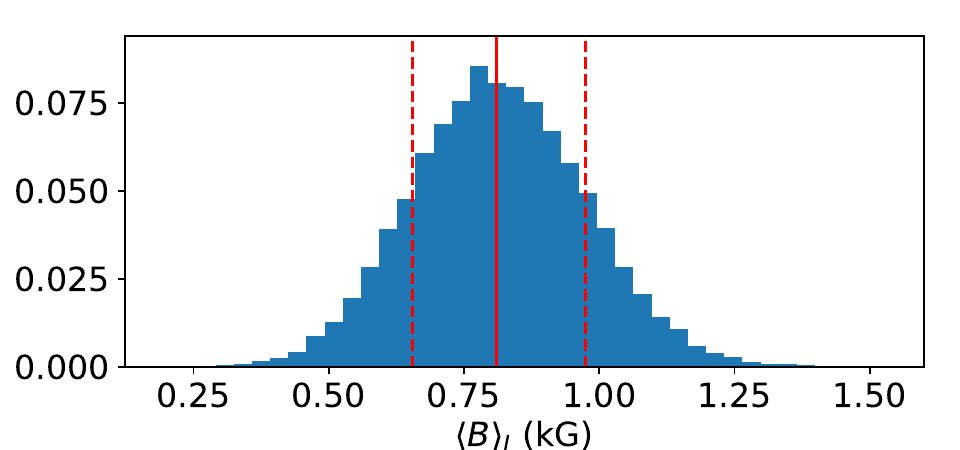}
    \caption{Magnetic field posterior distributions of Sample 2 of AK Sco, same as in Fig. \ref{fig:DQTauBdist}for DQ Tau. The top and bottom is for the A and B component
respectively}
    \label{fig:AKScoBdist}
\end{figure}

\begin{table}
    \centering
    \caption{Obtained magnetic field parameters for the components of AK Sco. See Fig. \ref{fig:AKScoSmallScale} for a complete list of parameters.}
    \begin{tabular}{l ll ll}
    \hline
     & \multicolumn{2}{l}{Sample 1} & \multicolumn{2}{l}{Sample 2} \\
    Parameter & A & B & A & B\\
    \hline
    $\langle B\rangle_I$ (kG) & $0.93\pm0.20$ & $1.05\pm0.22$ & $0.95\pm0.16$ & $0.81\pm0.16$ \\
    $f_{2}$ & $0.47\pm0.10$ & $0.52\pm0.11$ & $0.48\pm0.08$ & $0.41\pm0.08$ \\
    LR & \multicolumn{2}{l}{$1.06\pm0.02$} & \multicolumn{2}{l}{$0.96\pm0.01$} \\
    \hline
    \end{tabular}
    
    \label{tab:small-scaleAK}
\end{table}

\section{Discussion}
\label{sec:discuss}

\subsection{DQ Tau: a stable unstable accretion}

Even if our previous work on DQ Tau \citep[][Paper I]{Pouilly23} allowed us to enhance our understanding of the accretion process and the magnetic field of this system, a second set of observations, a few orbital cycles later, was needed to understand the implication of the binary motion and interactions on these parameters.
This new ESPaDOnS data set of 11 observations is divided over three orbital cycles with four observations on the first one, one on the second, and six on the last one. 
A $\sim$1-day sampling was achieved on the first and third cycles.

The orbital parameters derived from these observations are in favour of the apsidal motion detected in Paper I ($\dot{\omega}$ = 1.15 $\pm$ 0.25 $^\circ$yr$^{-1}$) with an increase of 1.7$^\circ$ in the timespan of about 1.75 years.
However, we have to note that the expected motion during the two epochs falls into the uncertainties of the argument of periastron's derivation.

The suspected rotation period of the secondary around 4.5 days has also been recovered (estimated around 4.8 days in Paper I). 
The ZDI analysis converged on a period of 4.6 days for the secondary, consistent with the 2D periodograms of Paper I.
Furthermore, the 2D periodograms of Balmer and He~\textsc{I} D3 lines are exhibiting the same signal at the same velocities as in Paper I, suggesting that this is a real periodic variation of the secondary's lines.
However, because of the observation sampling, we had to use only the last six observations, yielding a high FAP preventing us from claiming a definite period detection on this data set.

During the orbital cycle studied in Paper I, both components were showing accretion signatures but the main accretor was the secondary, the only one showing a NC of the He~\textsc{I} D3 line tracing the accretion shock.
During observations, the accretion seems fairly distributed between the two components, more similar to the epochs studied by \cite{Fiorellino22}. 
Following the main accretor from H$\alpha$ main emission's velocity, the secondary is the main accretor at phase 0.02 and 0.23, and the primary is so from phases 0.77 to 0.98 (see Table~\ref{tab:mainAcc}).
Furthermore, the He~\textsc{I} D3 line is clearly showing a NC at the primary's velocity more frequently, confirming that the magnetospheric accretion is ongoing on this component as well.
Despite the clear accretion signatures for both components, their variability is lower and does not show any maximum at the apastron, only at the periastron.
However, the double-peaked shape of Balmer and He~\textsc{I} D3 lines, especially at $\phi$=0.92, as well as the higher mass accretion rate, indicate that the two components are accreting simultaneously at a similar rate.

The magnetic analysis helps to understand what changed between Paper I's observations and the present orbital cycle.
Even if the large-scale magnetic field of the secondary seems stable, the ZDI analysis revealed significant changes in the primary's topology.
Indeed, the magnetic field strength and the contribution of the dipolar component contribution to the total magnetic energy have slightly increased, but the strongest variation is the almost 25\% increase in the toroidal field contribution, now dominating the primary's magnetic topology.
As the accretion in DQ Tau seems magnetically driven, one can hypothesize that both the secondary's accretion and magnetic field are stable, while the primary is responsible for the inter-cycle variations of the accretion signatures.

Along the orbital cycles presented in this work, the magnetic field seems stable.
This is shown by the observations at HJD 2~459~835.01 and 2~459~868.10 ($\phi_{\rm orb}$ = 0.83 and 0.92, $\phi_{\rm rot}$(A) = 0.331 and 0.253, respectively), that are showing very similar Stokes I and V profiles for the primary, despite the 10 rotation cycles between them. 
For the B component, the observations at HJD 2~459~837.04, HJD 2~459~868.98, and HJD 2~459~872.94 ($\phi_{\rm orb}$ = 0.96, 0.98, and 0.23, $\phi_{\rm rot}$(B) = 0.347, 0.396, and 0.271) exhibit consistent behaviour, as for HJD 2~459~838.03 and HJD 2~459~869.95 ($\phi_{\rm orb}$ = 0.02 and 0.04, $\phi_{\rm rot}$(B) = 0.562 and 0.611).

The large-scale topology of the secondary's magnetic field behaves as expected for low-mass, fully convective, cTTS, showing a highly axi-symmetric field with a strong dipole component \citep{Gregory12, Villebrun19}.
Nevertheless, both components lie at the edge of the boundary of the bistable dynamo regime derived by \cite{Gregory12}, which seems to be the regime of the primary, on the weak field dynamo branch, with its less axi-symmetric and weaker magnetic field (when the secondary could be on this regime as well, but on the strong field dynamo branch).
This is reminiscent of V2247 Oph \citep{Donati10a}, which is a fully convective TTS hosting a complex magnetic field with a weak dipole component.
However, DQ Tau's components are more massive than V2247 Oph (at 0.35 \msun), the $\sim$0.6 \msun\ mass could thus represent the high-mass end of the bistable regime of cTTSs.

Unlike the large-scale field, it appears that the average small-scale fields have remained more stable on the two components, showing magnetic field strengths of 2.8 and 2.4 kG (c.f. 2.7 and 2.4 kG in Paper I) for the A and B components, respectively.
This is particularly interesting on the A component, where there has been a significant evolution of the large-scale field structure without much variation on the small-scale in the $\sim$ 2 years between the epochs.
Similarly to other studies of binary stars \citep[e.g.][]{Kochukhov19,Hahlin22,Pouilly23}, this points towards stronger stability of the small-scale fields.
In the case of DQ Tau, this is particularly interesting, as this is the first case where a binary has had magnetic field measurements on multiple spatial scales at several epochs.
This shows the stability of the small-scale fields over any cyclic activity that might be ongoing which causes the large-scale field to vary.

The last point to discuss is the accretion enhancement at the apastron.
Even if the origin of this phenomenon is not clear, the increase in the primary's accretion and the change of its magnetic topology are occurring together with the absence of this apastron enhancement of the mass accretion rate.
The magnetic variation of the primary might thus be at the origin of the cycle-to-cycle variation of the whole system's accretion process.
The increase of the toroidal field is reminiscent of the magnetospheric inflation scenario \citep{Bouvier03}, observed on several single cTTS \citep[e.g.,][]{Alencar18, Pouilly20, Pouilly21}.
In the single-star situation, a shear of the magnetic field lines due to a truncation radius different from the corotation radius induces a toroidal field which inflates the magnetic field lines up to an opening, before reconnection to a less sheared situation \citep{Zanni13}.
In the inflated stage, the mass accretion is lowered, enhanced at the reconnection, and a mass ejection called magnetospheric ejection occurs at the opening of the field lines.
It is difficult to extrapolate this phenomenon on binaries, but this is an indication that a link between toroidal field and mass accretion variation does exist.
In addition, or instead of this phenomenon, the magnetosphere of the two star separate and merge during every periastron passage, as evidenced by the magnetic reconnection events observed in the radio \citep{Salter10} and in X-rays \citep{Getman11, Getman22}. 

\subsection{AK Sco: episodic accretion and inter-cycle variations}

Only a few accreting spectroscopic binary systems such as DQ Tau are known, AK Sco is one of them.
Furthermore, this system is also orbiting on an eccentric orbit, with a similar period.
Studying this system is thus a unique opportunity to address the accretion processes of eccentric binaries, among the behaviours specific to each system.
The 19 HARPSpol observations we used consist of four different runs performed in June 2016, 2017, 2019, and April/May 2022.
Only the 2022 data set, containing 8 observations with a 1-day sampling, is suitable for detailed variability analysis as carried out for DQ Tau, but the well-known orbital motion, and the fairly good phase coverage of the orbital cycle by the four runs is enough to study the accretion-ejection signatures and draw a picture of how it is ongoing on AK Sco on the orbital timescale.

We thus focused our spectroscopic analysis on the only accretion-related line in emission of the AK Sco spectrum, H$\alpha$, which displays a large and complex variability we tried to disentangle.
H$\alpha$ mostly presents a double-peaked shape, already observed on cTTSs \citep[i.e.,][]{Bouvier03, Bouvier07a, Pouilly21} and characteristic of the magnetospheric accretion process.
Furthermore, the IPC profiles between phases 0.19 and 0.63, extending near the free-fall velocity and independently correlated over two velocity regions, are consistent with an accretion through funnel flows for the two components.
The mass accretion rate of about 10$^{-8.3}$ \msunyr\ is consistent with the typical values observed on cTTSs, and modulated on the orbital period, as expected for such accreting binary system.

Even if this looks like a stable, orbit-modulated, accretion through funnel flows forming at the apastron, the observation near the apastron were taken during the same orbital cycle, we thus do not have any proof that this phenomenon is stable, or occurring at each cycle.
Quite the contrary, we observed a clear variation of the ejection signatures between the cycle observed in 2016 and the one in 2017 at the same orbital phase (see Fig.~\ref{fig:haAKPeri}).
Furthermore, the TESS light curves taken in 2019 and 2021 are showing very different behaviours. 
The two 2019 sectors are showing a high frequency and low amplitude modulation while the two 2021 sectors seem modulated on a lower frequency, and with a higher amplitude.
The situation thus changed in a few years, in addition to the cycle-to-cycle changes, as shown by the minimum on the first 2021 cycle, which became a maximum on the second (around phase 0.6).
Such disparities are reminiscent of the inter-cycle variations of the accretion-ejection processes reported in the literature \citep{Alencar03, GomezDeCastro20}, as well as of the pulsed accretion observed on DQ Tau \citep[][and Paper I]{Tofflemire17, Fiorellino22}.

The magnetic analysis is quite challenging for this system.
The lack of Stokes V signatures is indicating a very weak large-scale magnetic field, or a very complex small-scale field inducing a large amount of polarities cancellation at large-scale, as expected given the very shallow or inexistent outer convective envelope of the components \citep{Alencar03, Gregory12, Villebrun19}.
We are thus not able to claim any definite detection of a large-scale magnetic field, nor compute ZDI maps as it was done for DQ Tau.
F-type stars are expected to be less magnetic than M-type stars , detecting their magnetic field is thus more challenging.
\cite{Villebrun19} reported a detection for such an object, CO~Ori (\teff~=~6290~K, $B_{\rm l}$~=~$-$96.7~G), with a LSD Stoke V profile showing SNR lower than most of our observations (7064), indicating that the observations' quality is not responsible of this lack of detection on AK~Sco.
However, the Fe~\textsc{I} multiplet around 550 nm, which contains both magnetically sensitive and insensitive lines, is suitable for a small-scale magnetic field analysis following the same Zeeman intensification procedure as DQ Tau.
The magnetic field strengths are similar and around 1 kG for both components (see Sect.~\ref{subsubsec:ZintensAK} for more details).
These smaller field strengths than DQ Tau are consistent with the absence of Stokes V signature, as the two spatial scales seem to be coupled \citep{Vidotto14, Hahlin23}, and even if this field strength might yield a large-scale field detection, we cannot exclude a complex topology yielding polarities cancellation to explain the lack of Stokes V signatures.
Furthermore, \cite{Hahlin23} revealed a disagreement between the small-scale field computed on Sun-like stars at optical and near-infrared wavelength, especially for the weaker field.
They concluded on a possible overestimation of the magnetic field strength when optical observations are used, which is the case of AK Sco in this work.

Even if \cite{GomezDeCastro13} showed that AK Sco is gravitationally attracting the material from its distant disk, the H$\alpha$ line shape and the IPC are in favour of a magnetically-driven accretion, probably when the material gets closer to the component.
The weakness \citep[according to][]{Jarvinen18} or lack of detection (according to our study) of global magnetic field is not critical for the magnetospheric accretion to occur, as previously reported in the case of HQ Tau and V807 Tau \citep[][respectively]{Pouilly20, Pouilly21}.

\subsection{Eccentric binaries: toward a common accretion scheme ?}

Here we address if there is, or is not, a global scheme of the accretion process on eccentric binaries, in addition to behaviours specific to each system.
The main advantage of studying these two systems, DQ Tau and AK Sco, is that they are very similar binary systems.
Their components are young, accreting material from a circumbinary disc, with similar masses, on an eccentric orbit with a period of a dozen days.
Of course, two objects are not enough to claim the characterisation of the eccentric binaries' accretion in general, but we can already address similarities in their processes which can be ascribed to the eccentric orbit.

The most evident is the modulation of the mass accretion rate on the orbital period.
This is due to the maxima observed at the periastron and the apastron on both objects, even if the apastron's enhancement does not occur at every orbital cycle.
These enhancements are similar in amplitude, of about an order of magnitude at periastron, and more moderate (half an order of magnitude) when they occur at apastron.

Concerning the preferred accreting component, DQ Tau seems to vary between the secondary as the principal accretor (see Paper I) and a more balanced situation as it is shown in this work, which is reflecting an evolution of the primary's large-scale magnetic field. 
This is in agreement with the finding of \cite{Fiorellino22}, who found that both stars are accreting and the principal accretor can change between the primary and the secondary over a few orbits.
AK Sco, during the cycles studied in this work, is showing the same behaviour as DQ Tau in Paper I, a study of another orbital cycle might reveal a balanced phase like DQ Tau.
In any case, the study of the main accretor through the velocity position in the main peak of H$\alpha$ is showing, for all orbital cycles studied up to now and for both systems, changes in the main accreting component (see Table \ref{tab:mainAcc}).

\begin{table}
    \centering
    \caption{Main accretor at each orbital phase, identified from the velocity position of the highest emission peak in H$\alpha$, H$\beta$, H$\gamma$,  and NC of He~\textsc{i} D3 lines for DQ Tau, and H$\alpha$ only for AK Sco. "A" and "B" indicate the primary or the secondary as the main accretor according to the given line, "?" indicates that we were not able to identify the main accretor from the given line at that phase.}
    \begin{tabular}{lccccclclc}
    \multicolumn{5}{c|}{DQ Tau} & \multicolumn{4}{|c}{AK Sco} \\
    \cmidrule(lr){1-5}\cmidrule(lr){6-9}
    $\phi_{\rm{orb}}$ & H$\alpha$ & H$\beta$ & H$\gamma$ &  He~\textsc{i} & $\phi_{\rm{orb}}$ & H$\alpha$ & $\phi_{\rm{orb}}$ & H$\alpha$ \\
    \cmidrule(lr){1-5}\cmidrule(lr){6-9}
    0.02 & B & A & B & A & 0.01 & A & 0.46 & B \\
    0.04 & B & B & B & A & 0.02 & A & 0.48 & B \\
    0.11 & B & A & A & A & 0.09 & ? & 0.54 & B \\
    0.17 & B & A & B & ? & 0.12 & A & 0.55 & B \\
    0.23 & B & B & B & A & 0.17 & ? & 0.62 & B\\
    0.42 & ? & ? & ? & ? & 0.19 & ? & 0.63 & B \\
    0.77 & A & B & A & ? & 0.26 & B & 0.94 & A \\
    0.83 & A & B & B & B & 0.33 & B & 0.95 & A \\
    0.92 & ? & ? & ? & A & 0.33 & B & & \\
    0.96 & A & B & B & ? & 0.39 & B & & \\
    0.98 & A & B & A & B & 0.40 & B & & \\
    \cmidrule(lr){1-5}\cmidrule(lr){6-9}

    \end{tabular}
    \label{tab:mainAcc}
\end{table}

The obvious inter-cycle variation of AK Sco's accretion-ejection processes through H$\alpha$ line (see Fig.~\ref{fig:haAKPeri}), seems to be a recurrent behaviour as well.
Indeed, by comparing the H$\alpha$ line of DQ Tau of this work and Paper I (see Fig.~\ref{fig:haPeri} and \ref{fig:ha01}), we noticed that the line shape and strength are varying significantly at similar orbital phases but different orbital cycles.
This is reminiscent of the primary's accretion enhancement observed on DQ Tau and thus might indicate that AK Sco is showing this kind of variation as well.
This inter-cycle variation is well seen on the two objects' light curves, even if the differences cycle-to-cycle are more moderate on DQ Tau (see \cite{Kospal18} for K2)
, significant changes in the periastron burst occur from one cycle to another.

Even if both systems are showing magnetospheric accretion signatures on their two components, the systems' magnetic fields are quite different, much weaker at small-scale and undetectable at large-scale on AK Sco.
The lack of Stokes V signatures prevents us to derive the magnetic topology of the components, but the spectroscopy is clearly showing that a magnetospheric accretion-like process is in place as well letting us suggest a dipole-dominated topology.
In any case, it seems that the magnetospheric accretion on eccentric binaries can occur over a wide range of large-scale magnetic field strengths.

\section{Conclusions}
\label{sec:ccl}

DQ Tau and AK Sco are two very similar systems composed of two equal-mass accreting PMS stars, on an eccentric orbit, and surrounded by a circumbinary disc.
Studying the accretion process of these systems and directly comparing the results might thus yield the establishment of a common accretion scheme for a young eccentric binary. 
As a key process of stellar evolution, understanding the accretion of young stellar objects is a crucial need, but still poorly studied for binary stars even if most low-mass Sun-like stars are born in multiple systems.
The study of DQ Tau itself is a follow-up of the work performed in Paper I, a few orbital cycles later, aiming to characterise the possible cycle-to-cycle variation regarding the accretion and the magnetic field of the system.
AK Sco being the closest, in terms of binary configuration, system to DQ Tau, comparing the two systems is thus the quickest way to start drawing the picture of the young eccentric binary's accretion.

The new orbital cycle of DQ Tau revealed a different accretion configuration, the primary is accreting more, enhancing the total mass accretion rate of the system, without particular enhancement at the apastron this time. 
Previously dominated by the secondary as the main accretor, the situation is more balanced during the orbital cycle studied in this work.
Furthermore, the magnetic topology of the A component significantly changed, and is now dominated by the toroidal field, which might be at the origin of the inter-cycle variations of the system's accretion.

AK Sco's components are showing a weaker magnetic field (around 1 kG at small-scale, and undetected at large scale) and a more moderate mass accretion rate.
Even though the system's accretion remains modulated at the orbital period and shows enhancement at the periastron as expected, the cycle-to-cycle variation of the accretion-ejection signatures is important.
Our spectroscopic measurements reveal signatures of magnetospheric accretion typical of single stars, even if the gravity plays a key role given the weak magnetic field, the distant inner edge of the circumbinary disc, and the higher mass of the components.

The orbit-modulation of the accretion is the more evident common behaviour of young eccentric binaries, as expected from the close separation at the periastron and the shorter distance to the disc at the apastron.
We reported cycle-to-cycle variations of the accretion process as well, which might be attributed to components' magnetic variations.
Finally, the magnetospheric accretion typical of single stars seems to occur in these systems as well, even if it is after a gravitational drag of the material coming from a distant inner disc.

\section*{Acknowledgements}

We thank the anonymous referee for the very pertinent comments provided, helping this article to be scientifically stronger and clearer.

Based on observations obtained at the Canada–France– Hawaii Telescope (CFHT) which is operated from the summit of Maunakea by the National Research Council of Canada, the institut National des Sciences de l’Univers of the Centre National de la Recherche Scientifique of France, and the University of Hawaii. The observations at the Canada–France–Hawaii Telescope were performed with care and respect from the summit of Maunakea which is a significant cultural and historic site.

Based on observations collected at the European Organisation for Astronomical Research in the Southern Hemisphere under ESO programmes 097.C-0227(A), 099.C-0081(A), 0103.C-0240(A), and 109.230J.001.

O.K. acknowledges support by the Swedish Research Council (grant agreement no. 2019-03548), the Swedish National Space Agency, and the Royal Swedish Academy of Sciences.

This project has received funding from the European Research Council (ERC) under the European Union's Horizon 2020 research and innovation programme under grant agreement No 716155 (SACCRED).

\section*{Data Availability}

The spectropolarimetric data used for DQ Tau in this work will be publicly released on 2024 February 28 at the Canadian Astronomy Data Center (CADC, \url{https://www.cadc-ccda.hia-iha.nrc-cnrc.gc.ca/en/}), program ID 22BF15.

The AK Sco observations used are available on the ESO HARPS archive (\url{https://www.ls.eso.org/sci/facilities/lasilla/instruments/harps_old/archive.html}), programs ID 097.C-0227(A), 099.C-0081(A), 0103.C-0240(A), and 109.230J.001.



\bibliographystyle{mnras}
\bibliography{DQTau_AKSco} 



\appendix
\section{Other DQ Tau's emission lines studied}
\label{ap:emlines}

In this section we present the two DQ Tau's Balmer residual lines studied in addition to the H$\alpha$ and He~\textsc{i} D3 lines (Fig.~\ref{fig:ha} and \ref{fig:he}).
These lines are described together with H$\alpha$ in Sect.~\ref{subsubsec:EmLineDQ}.

\begin{figure*}
    \centering
    \includegraphics[width=.45\textwidth]{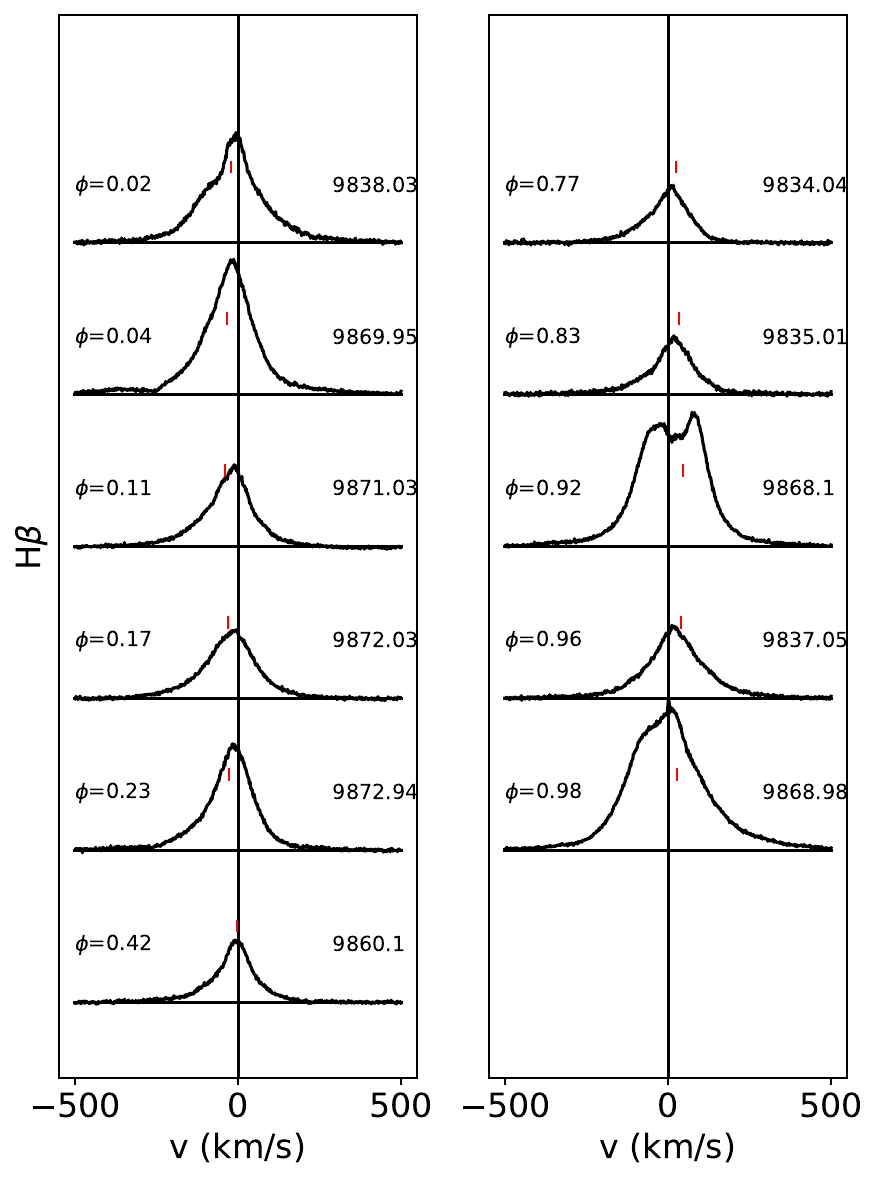}
    \includegraphics[width=.45\textwidth]{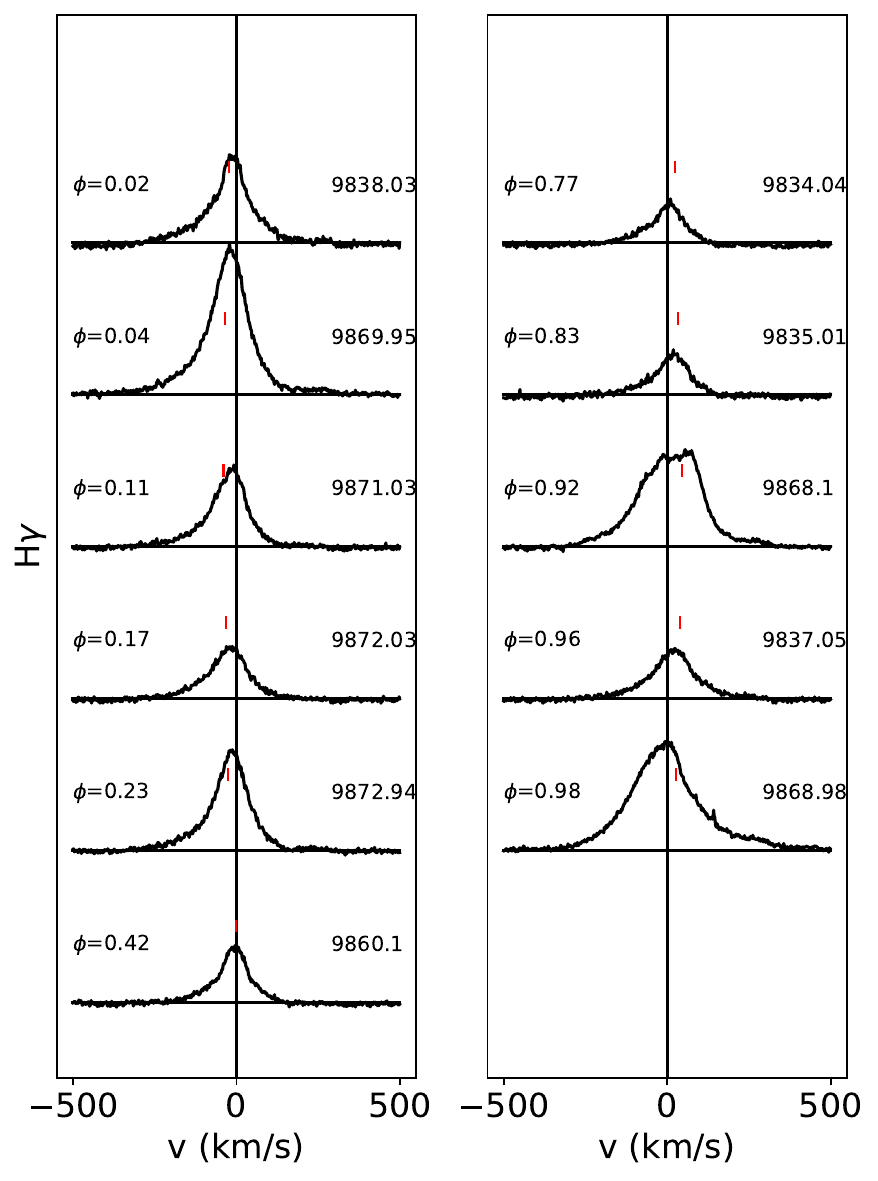}
    \caption{Same as Fig.~\ref{fig:ha} for DQ Tau's H$\beta$ (left) and H$\gamma$ (right) lines}
    \label{fig:hbhg}
\end{figure*}




\section{Inter-cycle variation in H$\alpha$ line}

\begin{figure*}
    \centering
    \includegraphics[width=.48\textwidth]{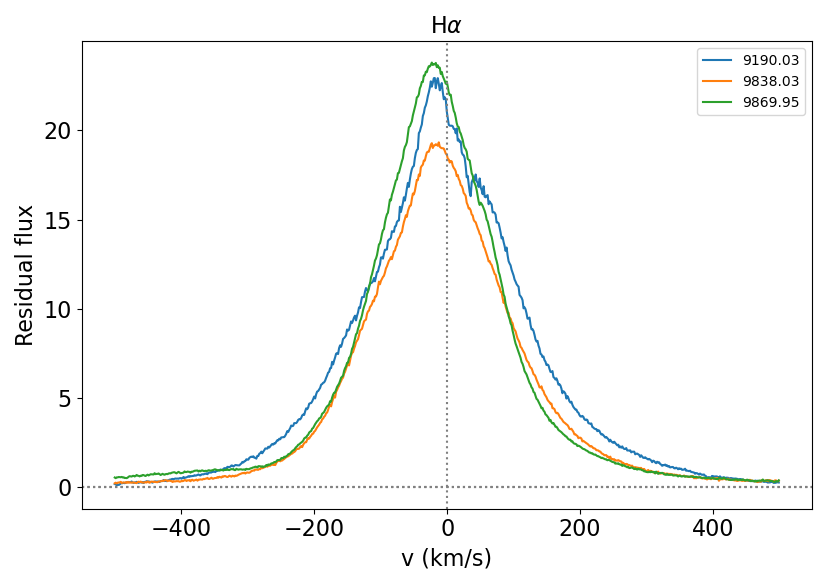}
    \includegraphics[width=.48\textwidth]{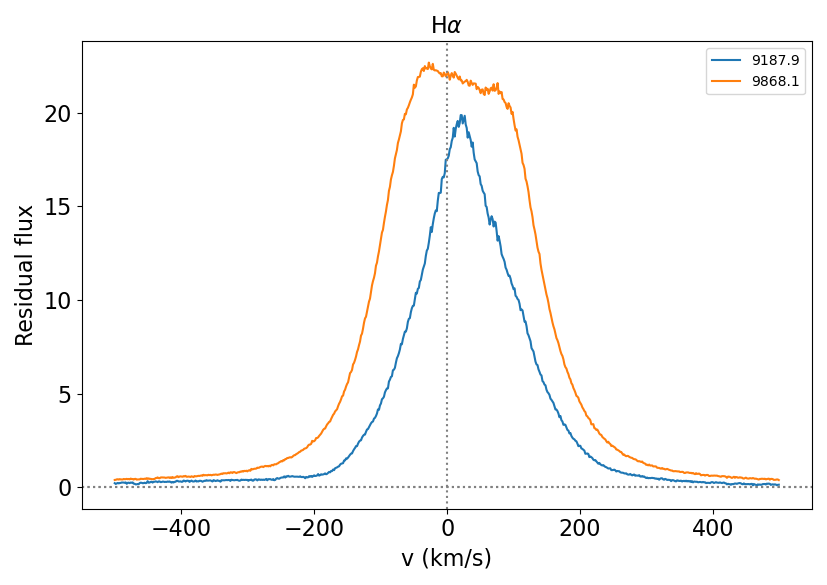}
    \caption{DQ Tau H$\alpha$ line. \textit{Left:} HJD 2~459~190.03 ($\phi_{orb}$=0.03), 2~459~838.03 ($\phi_{orb}$=0.02), and 2~459~869.95 ($\phi_{orb}$=0.04). \textit{Right:} HJD 2~459~187.9 ($\phi_{orb}$=0.9) and 2~459~868.1 ($\phi_{orb}$=0.92).}
    \label{fig:haPeri}
\end{figure*}

\begin{figure*}
    \centering
    \includegraphics[width=.48\textwidth]{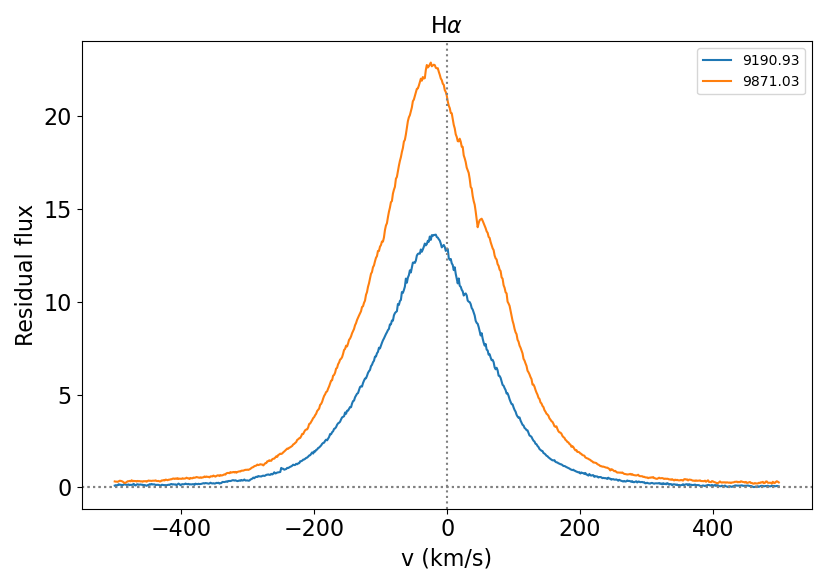}
    \includegraphics[width=.48\textwidth]{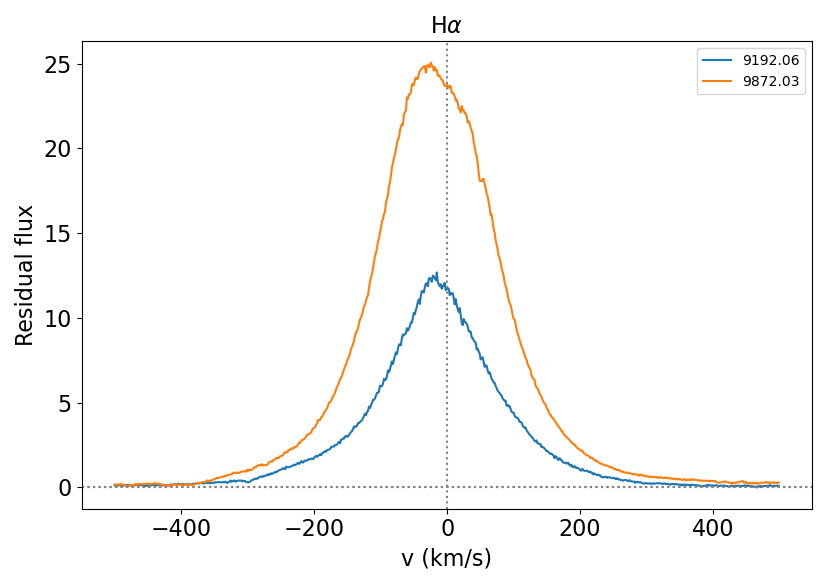}
    \caption{Same as Fig.~\ref{fig:haPeri}. \textit{Left:} HJD 2~459~190.93 ($\phi_{orb}$=0.09) and 2~459~871.03 ($\phi_{orb}$=0.11). \textit{Right:} HJD 2~459~192.06 ($\phi_{orb}$=0.16) and 2~459~872.03 ($\phi_{orb}$=0.17).}
    \label{fig:ha01}
\end{figure*}

\begin{figure*}
    \centering
    \includegraphics[width=.48\textwidth]{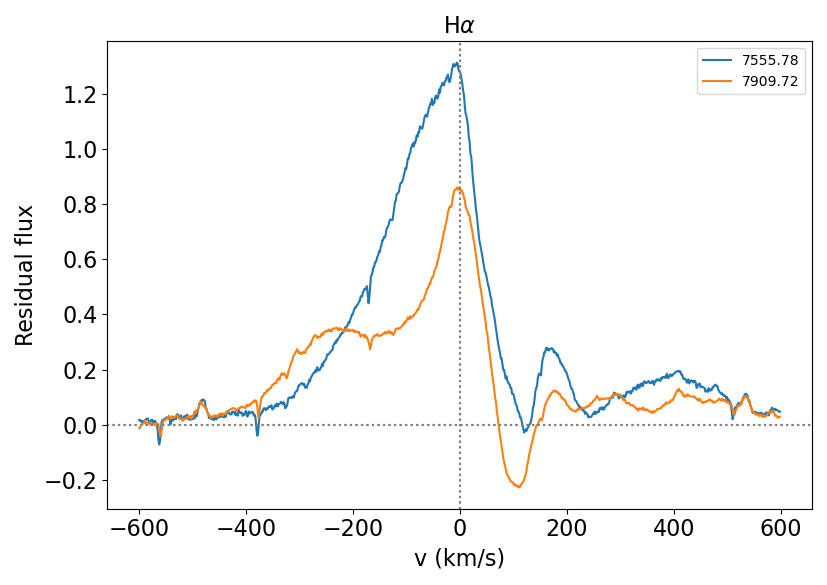}
    \includegraphics[width=.48\textwidth]{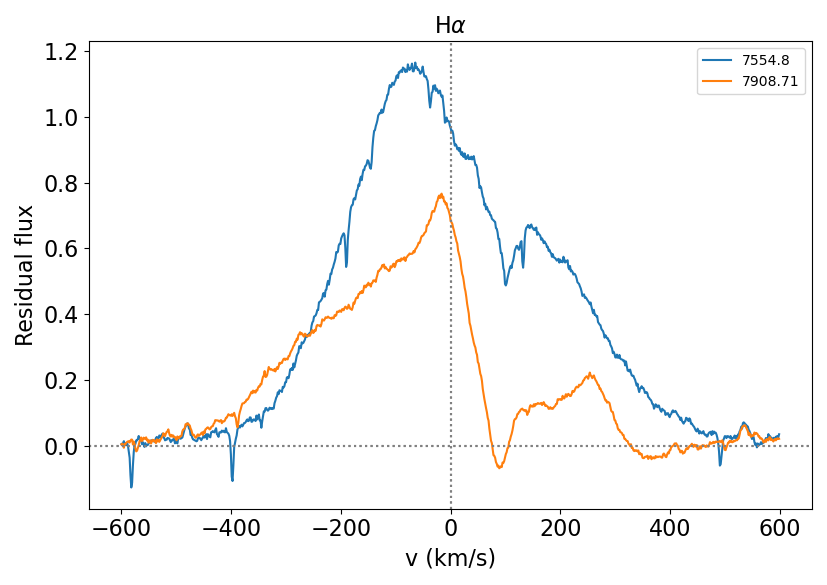}
    \caption{Same as Fig.~\ref{fig:haPeri} for AK Sco. \textit{Left:} HJD 2~457~555.78 ($\phi_{orb}$=0.01) and 2~457~909.72 ($\phi_{orb}$=0.02). \textit{Right:} HJD 2~457~554.8 ($\phi_{orb}$=0.94) and 2~457~908.71 ($\phi_{orb}$=0.95).}
    \label{fig:haAKPeri}
\end{figure*}

Here we present the evidence of inter-cycle variation of the H$\alpha$ lines.
Fig.~\ref{fig:haPeri} (Fig.~\ref{fig:ha01}) is showing the DQ Tau's lines around $\phi_{orb}$ = 0 ($\phi_{orb}$ = 0.1) at different orbital cycles.
The AK Sco lines around $\phi_{orb}$ = 0 are displayed in Fig.~\ref{fig:haAKPeri}.

\section{ZDI fit of LSD profiles}
\label{ap:fitZDI}

\begin{figure}
    \centering
    \includegraphics[width=.48\textwidth]{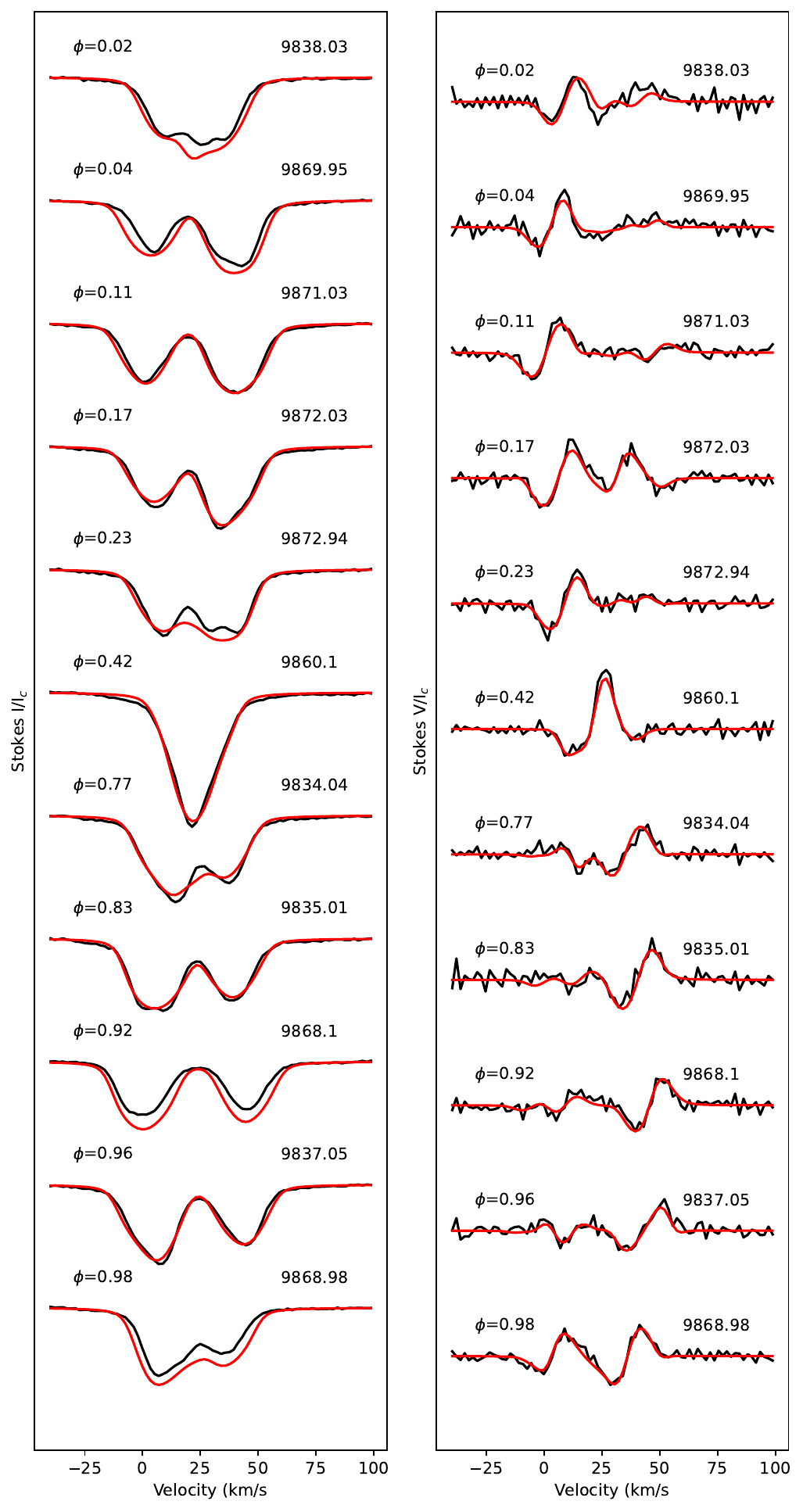}
    \caption{ZDI fit \textit{(red)} of DQ Tau LSD profile \textit{(black)}. The Stokes I profiles are on the left, while Stoke V profiles are on the right.}
    \label{fig:fitZDI}
\end{figure}

We are presenting in this section the results of the LSD Stokes I and V fit using ZDI for DQ Tau.
The fit are shown in Fig~\ref{fig:fitZDI} and the reduced $\chi^2$ are 5.27 on Stokes I and 1.33 on Stokes V.
The complete procedure is described in Sect.~\ref{subsec:ZDI}.

\section{Small-scale inference results}

\begin{figure*}
    \centering
    \includegraphics[width=\textwidth]{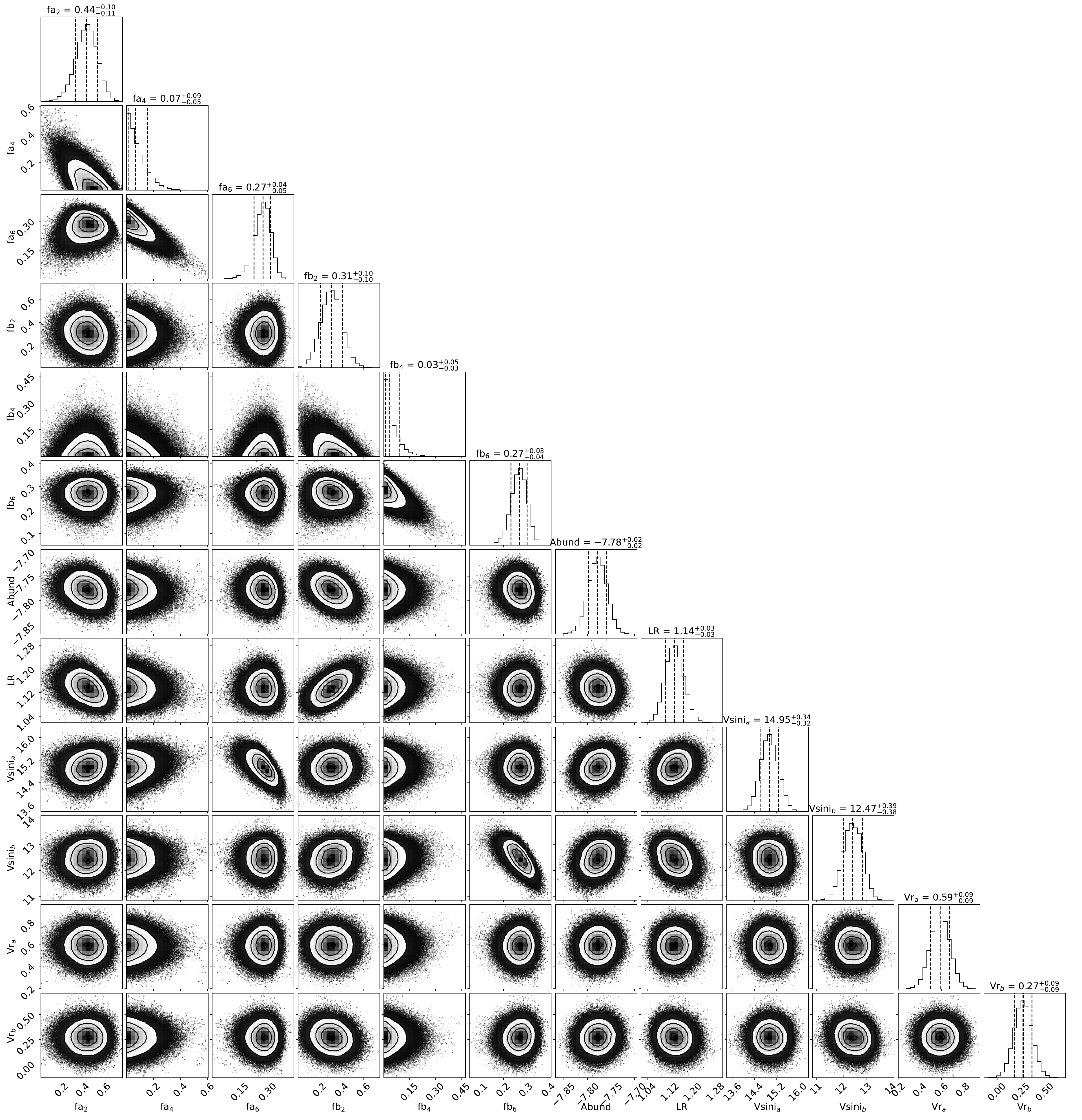}
    \caption{Corner plot showing the results of the magnetic field inference for DQ Tau.}
    \label{fig:DQTauSmallScale}
\end{figure*}

\begin{figure*}
    \centering
    \includegraphics[width=\textwidth]{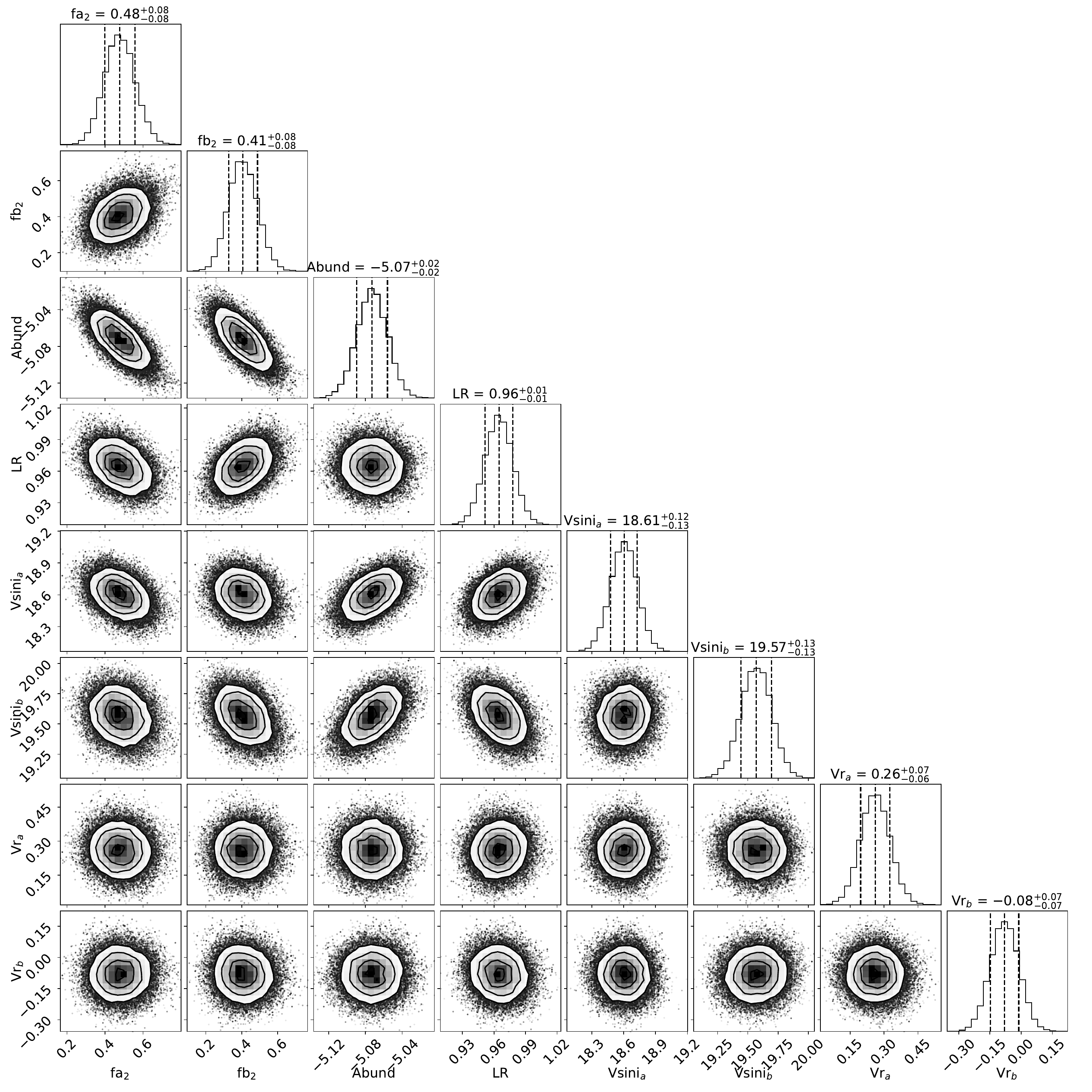}
    \caption{Corner plot showing the results of the magnetic field inference for Sample 2 of AK Sco.}
    \label{fig:AKScoSmallScale}
\end{figure*}

This section present the complete inference results of the Zeeman intensification analysis.
The corner plot of DQ Tau (Fig.~\ref{fig:DQTauSmallScale}) and AK Sco (Fig.~\ref{fig:AKScoSmallScale}) are resulting from the procedures described in Sect.~\ref{sec:small-scale} and \ref{subsubsec:ZintensAK}, respectively.

\bsp	
\label{lastpage}
\end{document}